\UseRawInputEncoding

\documentclass[11pt]{article}
\usepackage{longtable,supertabular}
\usepackage{amsmath}
\usepackage{amssymb}
\usepackage{latexsym}
\usepackage{cite}

\title{\bf List-decodable Codes and Covering Codes}
\author{Hao Chen
  \thanks{Hao Chen is with the College of Information Science and Technology/Cyber Security, Jinan University, Guangzhou, Guangdong Province, 510632, China, haochen@jnu.edu.cn. The research of Hao Chen was supported by NSFC Grant 62032009.}}

\begin{document}

\maketitle
\begin{abstract}
The list-decodable  code has been an active topic in theoretical computer science since the seminal papers of M. Sudan and V. Guruswami in 1997-1998. There are general results about the list-decodability to the Johnson radius and the list-decoding capacity theorem. However few results about general constraints on rates, list-decodable radius and list sizes for list-decodable codes have been obtained. List-decodable codes are also considered in rank-metric, subspace metric, cover-metric, pair metric and  insdel metric settings. In this paper we show that  rates, list-decodable radius and list sizes are closely related to the classical topic of covering codes. We prove new general simple but strong upper bounds for list-decodable codes in general finite metric spaces based on various covering codes.  The general covering code upper bounds can be applied to the case that the volumes of the balls depend on the centers, not only on the radius. Then any good upper bound on the covering radius or the size of covering code can be translated to a good upper bound on the sizes of list-decodable codes. Our results give exponential improvements on the recent generalized Singleton upper bound in STOC 2020 for Hamming metric list-decodable codes, when the code lengths are large.  A generalized Singleton upper bound for average-radius list-decodable codes is also given from our general covering code upper bound. We apply our general covering code upper bounds for list-decodable rank-metric codes, list-decodable subspace codes, list-decodable insertion codes and list-decodable deletion codes, list-decodable sum-rank-metric codes and list-decodable permutation codes. Some new better results about non-list-decodability of rank-metric codes, subspace codes,  sum-rank-metric codes and permutation codes with various metrics are obtained.\\
\end{abstract}

\section{Introduction}

For a vector ${\bf a} \in {\bf F}_q^n$, the Hamming weight $wt({\bf a})$ of ${\bf a}$ is the number of non-zero coordinate positions. The Hamming distance $d_H({\bf a}, {\bf b})$ between two vectors ${\bf a}$ and ${\bf b}$ is defined to be the Hamming weight of ${\bf a}-{\bf b}$. For a (linear) code ${\bf C} \subset {\bf F}_q^n$ of dimension $k$, its Hamming distance (or weight) $d_H$ is the minimum of Hamming distances $d_H({\bf a}, {\bf b})$ between any two different codewords ${\bf a}$ and ${\bf b}$ in ${\bf C}$. It is well-known that the Hamming distance (or weight) of a linear code ${\bf C}$ is the minimum Hamming weight of its non-zero codewords. The famous Singleton bound $d_H \leq n-k+1$, see \cite{Singleton},  is the basic upper bound for linear error-correcting codes. A linear code attaining this bound is called a MDS (maximal distance separable) code. \\

For a code ${\bf C} \subset {\bf F}_q^n$, we define its covering radius  by $$R_{covering}({\bf C})=\max_{{\bf x} \in {\bf F}_q^n} \min_{{\bf c} \in {\bf C}} \{wt({\bf x}-{\bf c})\}.$$ Hence the Hamming balls $B(x, R_{covering}({\bf C}))$ centered at all codewords $x \in {\bf C}$, with the radius $R_{covering}({\bf  C})$ cover the whole space ${\bf F}_q^n$.  We refer to the excellent book \cite{CHLL} on this classical topic of coding theory.  Actually the covering radius $R_{covering}({\bf C})$ of a linear $[n, k]_q$ code ${\bf C} \subset {\bf F}_q^n$ can be determined as follows. If $H$ is any $(n-k) \times n$ parity check matrix of ${\bf C}$, $R_{covering}({\bf C})$ is the least integer such that every vector in ${\bf F}_q^{n-k}$ can be represented as ${\bf F}_q$ linear combinations of $R_{covering}({\bf C})$ or fewer columns of $H$, see \cite{Janwa1,DMP}. It follows the redundancy upper bound $$R_{covering}({\bf C}) \leq n-k$$ for a linear $[n, k]_q$ code, see \cite{CHLL}. Let $n$ be a fixed positive integer and $q$ be a fixed prime power, for a given positive integer $R<n$, we denote $K_q(n, R)$ the minimal size of a code ${\bf C} \subset {\bf F}_q^n$ with the covering radius smaller than or equal to $R$. Set $ \frac{log_q K_q(n, \rho n)}{n}=k_n(q, \rho)$. The following asymptotic bound is well-known,  $$1-H_q(\rho) \leq k_n(q,\rho) \leq 1-H_q(\rho)+O(\frac{logn}{n}),$$ where $H_q(r)=rlog_q(q-1)-rlog_q r-(1-r)log_q (1-r)$ is the $q$-ary entropy function, see Chapter 12 of \cite{CHLL} and \cite{CF}.  In particular when each vector in ${\bf F}_q^n$ is in exactly one Hamming ball centered in codewords in ${\bf C}$ with the radius of $R$, we call this code a perfect codes. The existence and determination of perfect codes is is a fascinating topic in coding theory related to many other topics of mathematics. Hamming code and Golay code are basic examples of perfect codes, see Chapter 11 of \cite{CHLL}.\\

Let ${\bf F}_q$ be an arbitrary finite field, $P_1,\ldots,P_n$ be $n \leq q$ elements in ${\bf F}_q$. The Reed-Solomon codes $RS(n,k)$ is defined by $$RS(n,k)=\{(f(P_1),\ldots,f(P_n)): f \in {\bf F}_q[x],\deg(f) \leq k-1\}.$$ This is a $[n,k,n-k+1]_q$ linear MDS codes from the fact that a degree $\deg(f) \leq k-1$ polynomial has at most $k-1$ roots. It is well-known that the covering radius of the Reed-Solomon code we have its covering radius $n-k$ if $n \leq q$, see \cite{CHLL}.\\

Reed-Solomon codes can be generalized to algebraic-geometric codes as follows. Let ${\bf X}$ be an absolutely irreducible  non-singular genus $g$ curve defined over ${\bf F}_q$. Let $P_1,\ldots,P_n$ be $n$ distinct rational points of ${\bf X}$ over ${\bf F}_q$. Let ${\bf G}$ be a rational divisor over ${\bf F}_q$ of degree $\deg({\bf G})$ satisfying $2g-2 <\deg({\bf G})<n$ and $$support({\bf G}) \bigcap {\bf P}=\emptyset.$$ Let ${\bf L}({\bf G})$ be the function space associated with the divisor ${\bf G}$. The algebraic-geometric code associated with ${\bf G}$, $P_1,\ldots,P_n$ is defined by $${\bf C}(P_1,\ldots,P_n, {\bf G}, {\bf X})=\{(f(P_1),\ldots,f(P_n)): f \in {\bf L}({\bf G})\}.$$ The dimension of this code is $$k=\deg({\bf G})-g+1$$ follows from the Riemann-Roch Theorem. The minimum Hamming distance is $$d_H \geq n-\deg({\bf G}).$$ The Reed-Solomon codes are just the algebraic-geometric codes over the genus $0$ curve. One achievement of the theory of algebraic-geometric codes is the sequence of algebraic-geometric codes over ${\bf F}_{q^2}$ satisfying the Tsfasman-Vl\'{a}dut-Zink bound $$R+\delta \geq 1-\frac{1}{q-1},$$ which is exceeding the Gilbert-Varshamov bound when $q \geq 7$.  We refer to \cite{Garcia,TV} for the detail. Reed-Solomon codes and algebraic geometry codes are the basic examples in the theory of list-decoding and list decodable codes.\\

A length $n$ code ${\bf C} \subset {\bf F}_q^n$ is called (combinatorial) $(d_{list},L)$ list-decodable if  the Hamming ball of the radius $d_{list}$ centered at any ${\bf x} \in {\bf F}_q^n$ contains at most $L$ codewords of ${\bf C}$.  Since the classical papers of P. Elias and J. M. Wozencraft \cite{Elias57,Elias91,Wozencraft}, the list-decoding has attracted some further research in \cite{B,ZP}. The Johnson bound claims that any length $n$ code over ${\bf F}_q$ with the minimum Hamming distance $\delta n$ is $((1-\sqrt{1-\delta})n, qn^2\delta)$ list-decodable, see \cite{Johnson,GRS}.  The 1997 paper of M. Sudan \cite{Sudan} gives a beautiful list-decoding algorithm for the Reed-Solomon codes with the rate less or equal to $\frac{1}{3}$ beyond half the minimum distances (attaining $1-\sqrt{2R}$). Then an improved list-decoding algorithm matching the Johnson radius $1-\sqrt{R}$ were given for  Reed-Solomon codes and algebraic-geometric codes in \cite{GS,GS1}.   The list-decodablity of Reed-Solomon codes beyond the Johnson radius efficiently or combinatorially has been a central question in the theory. For the recent progress we refer to \cite{BKR,RW14, ShangguanTamo}.\\

For random codes, the list-decoding capacity theorem asserts that for any positive real number $\epsilon>0$, there exists codes ${\bf C} \subset {\bf F}_q^n$ with the rate $R({\bf C}) \geq 1-H_q(r)-\epsilon$ such that ${\bf C}$ is $(rn, \frac{1}{\epsilon})$ list-decodable. If the rate is bigger than $1-H_q(r)+\epsilon$, the code is list-decodable with the exponential (in $n$) list size, we refer to \cite{Elias91,ZP}, Theorem 2.1 and Theorem 2.2 in \cite{Rudra}. For code families achieving the list-decoding capacity we refer to \cite{GR1,KRSW}. The existence of Reed-Solomon codes of the rate $\Omega(\epsilon)$, which are $(1-\epsilon, O(\frac{1}{\epsilon})$ list-decodable, was proved in \cite{FKS}, we refer to \cite{GLST} for the other method. For other results about the list-decodablity of Reed-Solomon codes beyond the Johnson radius, we also refer to \cite{GR,GHS,GW,GX,GX1,SW,Wootters}.\\

It is obvious that list-decodable codes can be considered  in other finite metric spaces. For some works on list-decoding and list-decodable codes in rank-metric spaces,  subspace metric spaces, cover metric spaces, pair metric spaces, and list-decodable insertion-deletion codes, we refer to \cite{GX1,GW1,WZ,RST,YDing,BS,RZ,SZ,MV13, WZ1,WZ2,LXY,LXY1,MV19,HSS18,HS20,HY20,GHS,LTX}.\\

\section{Related works and our contribution}

\subsection{Related works (\cite{JH01,GHSZ,GV05,GV10,GN,GLMRSW20})}

 Though there have been active research on list-decodable codes since 1997, very few general bounds on finite length list-decodable codes have been obtained. We refer to papers \cite{JH01, GHSZ,GV05,GV10,YP,GN}  for the asymptotical combinatorial bounds of list decodable codes  and the recent papers \cite{ShangguanTamo,GLMRSW20}.  The classical Singleton bound $$|{\bf C}| \leq q^{n-2d_{list}}$$ for the $(d_{list},1)$ list-decodable codes in \cite{Singleton} was generalized to $$|{\bf C}| \leq Lq^{n-\lfloor \frac{(L+1)d_{list}}{L}\rfloor}$$ for  $(d_{list}, L)$ list-decodable codes in the recent paper \cite{ShangguanTamo}. Several existence results about families of optimal list-decodable Reed-Solomon codes beyond the Johnson radius with the list sizes $L=2$ or $L=3$ were proved and in the case $L=2$, some explicit Reed-Solomon codes list-decaodable beyond the Johnson radius were given. It was conjectured that the generalized Singleton bound is tight for the Reed-Solomon codes over large enough fields in \cite{ShangguanTamo}. Hence there are indeed optimal Reed-Solomon codes list-decodable beyond the Johnson radius with the constant list sizes of  $2$ and $3$, attaining the generalized Singleton bound in \cite{ShangguanTamo}.\\

On the other hand some combinatorial bounds for list-decoding have been studied in \cite{JH01,GHSZ}.  Let ${\bf Z}^{+}$ be the set of positive integers and $l: {\bf Z}^{+} \longrightarrow {\bf Z}^{+}$ be a function of the list size. As in \cite{GHSZ},  $$radius({\bf C} , l)= \max \{e| \forall x \in {\bf F}_q^n, |B(x,e) \bigcap {\bf C}| \leq l\}, $$ $$Rad({\bf C}_i, l)=\inf \{\frac{radius({\bf C}_i, l(n_i))}{n_i}\}. $$ In the case $q=2$, for a given rate $R$, set $$U_l(R)=sup_{{\bf C}, R({\bf C}) \geq R} Rad({\bf C}_i, l),$$ $U_c^{const}$ and $U_c^{poly}$ mean the maximal of $U_l(R)$ for constant list size $c$ and the polynomial list size $c_1n^c$, $U^{const}(R)=\lim\sup_{c \longrightarrow \infty} U_c^{const}(R)$, $U^{poly}(R)=\lim \sup_{c \longrightarrow \infty} U_c^{poly}(R)$. One main result in \cite{GHSZ} is $$H^{-1}(1-\frac{1}{c}-R) \leq U_c^{const}(R) \leq H^{-1}(1-R).$$  The equality  $$U^{const}(R)=U^{poly}(R)=H^{-1}(1-R),$$ was proved in \cite{ZP},  where $H(x)$ is $2$-ary entropy function. The quantity $U_q^{const}(R)$ and $U_q^{poly}(R)$ can be defined similarly for any fixed prime power $q$. \\

Similarly for $q=2$ set $$L_l(\delta)=inf_{{\bf C}, relative-distance({\bf C}) \geq \delta} Rad({\bf C}_i,l)$$ is defined in \cite{GHSZ}.  $L_c^{const}(\delta)$ is the $L_l(\delta)$ for the constant list size $l(n)=c$, $L_c^{poly}(\delta) =\sup_{c_1} L_l(\delta)$ for $l(n) \leq c_1n^c$, $L^{const}(\delta)=\lim\sup_{c \longrightarrow \infty} L_c^{const}(\delta)$, $L^{poly}(\delta)=\lim \sup_{c \longrightarrow \infty} L_c^{poly}(\delta)$. Another main result in \cite{GHSZ} asserts  $$L_c^{poly}(\delta) <\delta,$$ and $$L^{poly}(\delta) \leq \frac{1}{2} (1-(1-2\delta)^{\frac{1}{2}+\epsilon}), $$ for $\delta=\frac{1}{2}(1-\Theta((log n)^{\epsilon-1})$. It was conjectured in \cite{GHSZ}  for $0 <\delta <\frac{1}{2}$, $$L^{const}(\delta)=L^{poly}(\delta)=\frac{1}{2}(1-\sqrt{1-2\delta}).$$ The quantity $L_q^{const}(\delta)$ and $L_q^{poly}(\delta)$ can be defined similarly for any fixed prime power $q$. \\

In \cite{YP} the problem of packing Hamming balls of the same radius in ${\bf F}_2^n$ with the constraint that these balls covering each point of ${\bf F}_2^n$ with the multiplicities at most $L$ is considered. Then asymptotic upper bounds on the list-decodable radius of binary codes improving the Blinovsky bound was given. For list-decodable binary codes, we also refer to \cite{ABP,Guruswamilec}.\\

In \cite{GV05,GV10,GR11} the upper bound for the size of $q$-ary length $n$ list-decodable code ${\bf C}$ with the radius $(1-\frac{1}{q})(1-\epsilon)n$ was given, it was showed that $|{\bf C}| \leq 2^{\frac{d_q}{\epsilon^2}log(\frac{1}{\epsilon})}$ for some constant $d_q$. In Corollary 4.5 by the using the covering radius of first order $q$-ary Reed-Muller code, the case with a varying $\epsilon_n$ going to zero is considered and a similar upper bound is given. In \cite{GN} the asymptotic upper bounds on the rate of average-radius list-decodable codes were given. A generalized Singleton upper bound for average-radius list-decodable codes is given in Subsection 4.5. In \cite{GLMRSW20} list sizes of random $(p,L)$ list-decodable and average-radius $(p,L)$ list-decodable codes with rate $1-H_q(p)-\epsilon$ were studied and bounded.\\

On the other hand there have been active research on list-decodability and list-decoding of rank-metric codes, subspace codes, cover-metric codes, pair metric spaces, insertion-deletion codes and sum-rank-metric codes, we refer to \cite{GX1,GW1,WZ,RST,WZ1,WZ2,BS,RZ,YDing,LXY,LXY1,MV13,MV19,SZ,HSS18,HY20,HS20,GHS,LTX,PR}. For example it was shown that Gabidulin codes and linearized Reed-Solomon codes in the sum-rank-metric can not be list-decodable beyond the Johnson radius, see \cite{WZ,RZ,YDing,PR}.  The constraint on list-decodable insertion-deletion code codes over general alphabets was given in \cite{HS20}. Efficient binary codes list-decodable to $1-\epsilon$ insertion and deletion errors was constructed in \cite{HSS18,GHS}. \\

\subsection{Our contribution}

The main contributions of this paper are as follows.\\

1) We look at list-decodable codes from the view of covering codes and then give covering code upper bound for the size of $(d,L)$ list-decodable codes in general finite metric spaces. This upper bound can apply to the case that the volumes of balls in this metric space ${\bf X}$ depend on centers. Then any good upper bound on the covering radius and any good upper bound on the size of covering codes with the radius $d$ would lead to a good upper bound on the size of arbitrary $(d,L)$ list-decodable codes. The covering-code point of view also suggests a probabilistic construction of list-decodable codes in general finite metric spaces.\\

2)  When the length is the polynomial of the field size, our covering code upper bounds give a lot of constraints if the upper bounds for covering codes are known in this range of length. For Hamming error-correcting code case, from many fascinating classic results about covering codes in \cite{CHLL}, many strong constraints such as Corollary 4.1, 4.5, 4.7 and 4.8 on list-decodable codes in Hamming metric spaces are presented.\\

3) In the Hamming metric setting, even for  $(\lfloor\frac{d-1}{2}\rfloor, 1)$ list-decodable codes, our covering code upper bounds give highly nontrivial upper bounds on the sizes of codes with the given minimum Hamming distances. For example from Corollary 4.8 for fixed $q\geq 8$ and $R$, if $t \geq 3(\lceil log_q+1\rceil+1)+2$, and $n \geq Rq^{(t-1)R}+2q^{t-2}+\Sigma_{j=3}^{\lceil log_q+1\rceil+3} q^{t-j}$, then a length $n$ $q$-ary code with the minimum Hamming distance $2R+1$ has at most $q^{n-tR}$ codewords. Our covering code upper bound gives many such upper bounds for codes with the minimum distance $d<<n$.\\

4) For Hamming error-correcting codes,  when code lengths are  large, exponential improvements on the generalized Singleton upper bound in the STOC 2020 paper \cite{ShangguanTamo} are given in Section 4. Actually in the case $L \geq d_{list}$ the generalized Singleton bound is just an easy corollary of our covering code upper bound plus the almost trivial redundancy upper bound $R_{covering} \leq n-k$ for linear $[n,k]_q$ codes. From the dual covering code the following upper bound for list-decodable codes is obtained.\\

{\bf Theorem 2.1.} {\em  Let $q$ be a fixed prime power and $m$ be a large positive integer. The code length $n$ satisfies  $(q^2-1)q^m \leq n \leq (q^2-1)q^{m-1}+2q$. For a given list-decodable radius $d_{list}$ and the above code length $n$,  let $u$ be the smallest positive integer such that $x(u,q,n) \leq d_{list}$. Then the cardinality of any $(d_{list}, L)$ list-decodable code ${\bf C} \subset {\bf F}_{q^2}^n$, satisfies $$|{\bf C}|\leq L \cdot q^{2u+\frac{n}{q-1}+\frac{n}{q^{\frac{m}{2}-2}}}.$$}\\

5) We give a generalized Singleton upper bound as follows (see Theorem 4.2) on the size of average-radius list-decodable codes in the Hamming metric setting, which is indeed stronger than the similar upper bound for list-decodable codes in some parameter range.\\

{\bf Theorem 2.2 (Generalized Singleton upper bound for average-radius list-decodable codes).} {\em Let $q$ be a prime power and $n$ be a positive integer satisfying $n \leq q$. Let ${\bf C} \subset {\bf F}_q^n$ be a $(d_{list}, L)$-average-radius list-decodable code then we have $|{\bf C}| \leq \max\{(L-1) (q^{n-d_{list}}-\displaystyle{n \choose d_{list}+1})+(\displaystyle{n \choose d_{list}-1}(q-1)^{d_{list}-1}+\displaystyle{n \choose d_{list}}(q-1)^{d_{list}})\displaystyle{n \choose d_{list}+1}, (L-2)q^{n-d_{list}}\}.$}\\

6) We give a sufficient condition (see Corollary 5.1) as follows for the non-list-decodability of rank-metric codes and constant dimension subspace codes, from which the non-list-decodability rooted in the sizes of the codes, not depending on other properties. Such kinds results for some Gabidulin codes and lifted Gubidulin codes,  which can not be list-decodable to any positive radius,  were proved in \cite{WZ,RZ,RST}.\\

{\bf Theorem 2.3} {\em Let $k, \rho, n$ be three positive integers satisfying $\rho <n, k <n$ and $$k-(n-\rho)(1-\frac{\rho}{2n}) \geq c$$ for a fixed positive real number $c$. If a rank-metric code ${\bf C} \subset {\bf M}_n({\bf F}_q)$ of the cardinality at least $q^{nk}$ is $(\rho, L)$ list-decodable, then the list size is at least $q^{cn}$,  exponential in $n$.}\\

7) An asymptotic sufficient condition (see Theorem 6.1) as follows on the rate for the non-list-decodability for subspace code in $Grass({\bf F}_q^n)$ of subspaces of all dimensions, with the subspace metric $d_S$ or the injection metric $d_I$, is given. To the best of our knowledge, there is no results about combinatorial list-decodability about these subspace code. For efficient list-decoding of these codes, we refer to the paper of Mahdavifar and Vardy \cite{MV19}.\\

{\bf Theorem 2.4.} {\em Asymptotically any $(rn, L)$ list-decodale code family in $(Grass({\bf F}_q^n), d_S)$ has its rate at most $1-2r$. Asymptotically any $(rn, L)$ list-decodale code family in $(Grass({\bf F}_q^n), d_I)$ has its rate at most $(1-2r)^2$.}\\

8) The concept of $(R_{list}, L_1, L_2)$ (combinatorial) covering list-decodable codes in general finite metric spaces is introduced. The (combinatorial) $(d_{list}, L)$ list-decodable codes are just $(d_{list}, L_1=0, L_2)$ covering list-decodable codes. We suggest to determine the $(R_{list}, L_1=L_2 \geq 1)$ covering codes. These codes can be thought as the generalization of perfect codes in the classical coding theory.\\

9) We give some new bounds on list-decodable  sum-rank-metric codes and list-decodable permutation codes with the Hamming metric and the Chebyshev metric.

\section{List-decodable codes and covering codes}

\subsection{Covering codes and covering list-decodable codes}

Let $({\bf X}, d)$ be a finite metric space, we assume that $d$ takes values in the set of non-negative integers. We say that a code ${\bf C} \subset {\bf X}$  has the covering radius $R_{covering}({\bf C})$, if the balls centered at all codewords with the radius $R_{covering}({\bf C})$ cover the whole space ${\bf X}$, and this $R_{covering}({\bf C})$ is the smallest radius with this property.  Let $K_{{\bf X}}(r)$ be the minimum size of covering code on ${\bf X}$ with the radius $r$. For a sequence of finite metric spaces $\{({\bf X}_n, d_n)\}$, and a fixed covering ratio $r$, we define $k(r)=\lim_{n \longrightarrow \infty} \frac{log_q(K_{{\bf X}_n}(rn))}{log_q(|{\bf X}_n|)}.$ Here $q$ is a prime power depending on the metric setting, that is, the code rate is defined as $rate({\bf C})=\frac{log_q|{\bf C}|}{log_q|{\bf X}|}$. This parameter is important to derive the asymptotic bound of non-list-decodability. \\

Let $B({\bf x}, R)$ be the ball centered at ${\bf x} \in {\bf X}$ with the radius $R$. For a covering code ${\bf C} \subset {\bf X}$ of the finite metric space $({\bf X}, d)$ with the covering radius $R$, we define the multiplicity at an arbitrary element ${\bf x} \in {\bf X}$ as $mul({\bf x}, {\bf C})=|\{{\bf c} \in {\bf C}: {\bf x} \in B({\bf c}, R)\}|$. Then the covering multiplicity of this covering code ${\bf C}$, $$mul({\bf C}) =\max_{{\bf x} \in {\bf X}} \{mul({\bf x}, {\bf C})\}.$$ We refer to \cite{CS} for the covering multiplicity of lattices.\\

Notice that in insertion or deletion error-correcting setting studied in \cite{WZ1,HY20,HSS18,HS20,GHS,LTX}, shorter or longer received words with different lengths are allowed in their definition of list-decodable codes. Hence we need to use the finite metric spaces of strings with different lengths.\\

A code ${\bf C} \subset {\bf X}$ is called (combinatorial) $(d,L)$ list-decodable if each ball centered at any element in ${\bf X}$ of the radius $d$ contains at most $L$ codewords of ${\bf C}$. Actually we have the following natural generalization of list-decodable codes.\\

{\bf Definition 3.1} {\em Let $({\bf X}, d)$ be a general finite metric space. A code ${\bf C} \subset {\bf X}$ is called $(R_{list}, L_1, L_2)$ covering list decodable if every ball in ${\bf X}$ of the radius $R_{list}$ contains at least $L_1$ codewords and at most $L_2$ codewords.}\\

An $(d_{list}, L)$ locally decodable code is an $(d_{list}, 0, L)$ covering list-decodable code. A $(R_{list}, L_1\geq 1, L_2)$ covering list-decodable code ${\bf C}$ has its covering radius $R_{covering}({\bf C}) \leq R_{list}$. A perfect code of minimum distance $d$ is a $(\frac{d-1}{2}, 1, 1)$ covering list-decodable code. Hence it would be interesting to study $(R_{list}, L_1=L_2)$ covering list-decodable codes For such codes, each ball of radius $R_{list}$ contains exactly $L_1=L_2$ codewords. These codes can be thought as generalized perfect codes.  For diameter perfect constant weight codes, see \cite{Etzion21}. We refer to \cite{CHLL} for previous generalizations of perfect codes.\\

For a covering code with the covering radius $R_{covering}({\bf C})$ and the multiplicity $mul({\bf C})$, this code is an $(R_{covering}({\bf C}), 1, mul({\bf C}))$ covering list-decodable code. Hence in the Hamming metric setting, it is interesting to ask the following question. Given fixed positive real number $r<1$, is there a family of covering codes $\{{\bf C}_n\}$ of ${\bf F}_q^n$ with the covering radius $R_{covering}({\bf C}_n)=rn$ and the multiplicity $mul({\bf C}_n) \leq poly(n)$?\\

For the view of efficient decoding, if we consider any covering code ${\bf C}$ as an $(R_{covering}({\bf C}), 1, L)$ covering list-decodable code, it seems not trivial to ask an efficient decoding for some well-structured covering list-decodable codes.  For example for a perfect code ${\bf C}$ in Hamming metric space as an $(\frac{d_H({\bf C})-1}{2}, L_1=L_2=1)$ covering list-decodable code, the decoding to the covering radius is just the unique decoding up to $\frac{d_H({\bf C})-1}{2}$.\\

\subsection{Covering code upper bounds}

The main result of this paper is the following covering code upper bound for the list-decodable codes in general finite metric spaces.\\

{\bf Theorem 3.1 (General covering code upper bounds).} {\em 1)Let $({\bf X}, d)$ be general finite metric space. Let ${\bf C}  \subset {\bf X}$ be an  $(d, L)$ list-decodable code. Suppose that ${\bf C}' \subset {\bf X}$ is a code with the covering radius $R_{covering} \leq d$, then we have $$|{\bf C}| \leq L|{\bf C}'|.$$  Moreover if ${\bf C}_1 \subset {\bf X}$ is an $(R, L_1 \geq 1, L_2)$ covering list-decodable code. Let ${\bf C}'' \subset {\bf X}$ be a code of the minimum Hamming distance $d({\bf C}'') \geq 2R+1$. and ${\bf C}'$ be a covering code with the radius $R$. Then $$L_1 |{\bf C}''| \leq |{\bf C}_1| \leq L_2 |{\bf C}'|.$$\\
2) Asymptotically for any fixed small positive $\epsilon$, if a code family ${\bf C}_i, i=1,2,\ldots,$ of $rate({\bf C}_i) \geq k(r)+\epsilon$ is $(rn,L_n)$ list-decodable, the the list size $L_n$ has to be exponential in $n$.}\\

{\bf Proof.} This covering code upper bound is almost obvious. The balls of the radius $d \geq R_{covering}({\bf C}')$ centered at the codewords of ${\bf C}'$ cover the whole space ${\bf X}$. Then in each such ball there are at most $L$ codewords of ${\bf C}$. We have $|{\bf C}| \leq L \cdot |{\bf C}'|$. The other conclusions follow directly.\\

This  covering code upper bound is strong since we can take any code ${\bf C}'$ with the covering radius $R_{covering}({\bf C}) \leq d$. Actually any good upper bound for the covering radius implies a good upper bound on the size of $(d, L)$ list-decodable codes from our covering code upper bound.  For a given code ${\bf C}$, we also can use our covering code bounds to lower bound the list sizes if ${\bf C}$ is $(d, L)$ list decodable.\\

\subsection{A probabilistic construction of list-decodable codes from good covering codes}

Let $({\bf X}, d)$ be a finite metric space satisfying the condition that the cardinalities of balls with the same radius $r$ are the same $B(r)$. Let ${\bf C} \subset {\bf X}$ be a covering code with the radius $R$.  Let $multi({\bf C}, {\bf X})$ be the set of points in ${\bf X}$ contained in at least two balls centered at two different codewords of ${\bf C}$ with the radius $R$, or the set of ${\bf x} \in {\bf X}$ satisfying $mul({\bf x}, {\bf C}) \geq 2$. Set $multiratio({\bf C}, {\bf X})=\frac{|multi({\bf C}, {\bf X})|}{|{\bf X}|}$. If this multiratio is very small, this covering code is good. For example if the covering code is perfect, that is, ${\bf X}$ is the disjoint union of balls centered at codewords of ${\bf C}$ with the radius $R$, the multiratio is zero.\\

Then we give a probabilistic construction of list-decodable codes from nice covering codes. For each ball of radius $R$ centered at a codeword of ${\bf C}$, $L$ points of ${\bf X}$ are chosen uniformly in this ball. Then the set of all these uniformly chosen points is a list-decodable code ${\bf C}'$ with high probability when the $multiratio({\bf C}, {\bf X})$ is very small. It is obvious that there are at least $(1-\frac{multiratio({\bf C}, {\bf X})} {B(R)})L \cdot |{\bf C}|$ and at most $L \cdot |{\bf C}|$  codewords in this code ${\bf C}'$.\\

{\bf Theorem 3.2.} {\em The above code ${\bf C}'$ is an $(R, L(1+\frac{multiratio({\bf C}, {\bf X})} {B(R)}))$ list-decodable code with a high probability.}\\

Interestingly we observe that when $muitiratio({\bf C}. {\bf X}) <<1$ is much smaller than 1, then the set of all points ${\bf x} \in {\bf X}$ satisfying $mul({\bf x},  {\bf C}) \geq 2$ is very few. That is, for most received words ${\bf x} \in {\bf X}$, there is only one ball centered in the codeword of ${\bf C}$  with the radius $R_{covering}({\bf C})$ contains ${\bf x}$. This means for most received words the list-decoding of ${\bf C}$ to the radius $R_{covering}({\bf C})$ degenerates to the unique decoding. Hence it would be interesting to construct and decode covering list-decodable codes with such very small $multiratio({\bf C}, {\bf X})$. We refer to \cite{Chen1}.\\

\section{Hamming metric}

\subsection{Covering code upper bounds for Hamming error-correcting codes}

From Theorem 3.1 we have the following result for $(d_{list},L)$ list-decodable codes in Hamming metric setting.\\

{\bf Theorem 4.1 (Covering code upper bounds for Hamming error-correcting codes).} {\em Let ${\bf C} \subset {\bf F}_q^n$ be an  $(d_{list}, L)$ list-decodable code. Suppose that ${\bf C}' \subset {\bf F}_q^n$ is a code with the covering radius $R_{covering} \leq d_{list}$, then we have $$|{\bf C}| \leq L|{\bf C}'|.$$ Hence we have $$|{\bf C}| \leq L K_q(n,d_{list}).$$ Moreover if ${\bf C}_1 \subset {\bf F}_q^n$ is an $(R_{list}, L_1 \geq 1, L_2)$ covering list-decodable code.  Then $$K_q(n, R_{list}) \leq |{\bf C}_1| \leq L_2|{\bf C}'|.$$  Let ${\bf C}'' \subset {\bf F}_q^n$ be a code of the minimum Hamming distance $d_H({\bf C}'') \geq 2R_{list}+1$. Then $$L_1 |{\bf C}''| \leq |{\bf C}_1| \leq L_2 |{\bf C}'|.$$}.\\

The covering upper bounds for a $(d_{list}, L)$ list-decodable code ${\bf C} \subset {\bf F}_q^n$ are weaker than the sphere-packing upper bound$$|{\bf C}| \leq L \cdot \frac{q^n}{\Sigma_{j=0}^{d_{list}} \displaystyle{n \choose j}(q-1)^j}.$$ This sphere-packing upper bound was formulated in \cite{Kopparty} for binary codes. However when $n$ is large, it is computational infeasible to get clear and explicit upper bounds from this expression. Hence by using various results in covering codes, we can give explicit upper bounds for list-decodable codes in Hamming metric setting.\\

There are a lot of classical results about the binary covering codes, we refer to \cite{CHLL,Litsyn,DJ,BDMP}. From the table \cite{Litsyn} of S. Litsyn, we have the following upper bound on the length $16$ binary $(3, L)$ list decodable code ${\bf C} \subset {\bf F}_2^{16}$, $$|{\bf C}| \leq 192L,$$ since $K_2(16,3) \leq 192$.  Thus if the linear $[16, 9, 4]_2$ code is $(3, L)$ list-decodable, the list size has to be at least $3$.\\

The almost trivial redundancy upper bound $R_{covering} \leq n-k$ for the covering radius of the linear $[n,k]_q$  code, see page 217 of \cite{CHLL}, and our covering code upper bounds imply an upper bound which is close to the generalized Singleton upper bound in \cite{ShangguanTamo}.\\

{\bf Corollary 4.1 .} {\em Let ${\bf C}'$ be a linear $[n, k]_q$ code in ${\bf F}_q^n$, then $R_{covering}({\bf C}') \leq n-k$. Hence an $(d_{list},L)$ list-decodable code ${\bf C} \subset {\bf F}_q^n$ have to satisfy $$|{\bf C}| \leq Lq^{n-d_{list}}.$$}\\

{\bf Proof.} The first conclusion follows that for any $k$ information set coordinate positions $1 \leq i_1<i_2 \cdots <i_k\leq n$, the coordinates at these positions of codewords in ${\bf C}'$ can be arbitrary vectors in ${\bf F}_q^k$.  For the second conclusion, we take a linear $[n, n-d_{list}]_q$ code ${\bf C}'$, then $R_{covering}({\bf C}') \leq d_{list}$. The conclusion follows immediately.\\

This is weaker than the generalized Singleton upper bound $$|{\bf C}| \leq Lq^{n-\lfloor \frac{(L+1)d_{list}}{L}\rfloor}.$$  in \cite{ShangguanTamo}. However when the list size $L \geq d_{list}$, this almost trivial upper bound from our covering code upper bounds plus the redundancy upper bound for the covering radius is equivalent to the generalized Singleton upper bound in \cite{ShangguanTamo}. We should mention that the covering radius of a $[n,k]_q$ Reed-Solomon code with the length $n \leq q$ is $n-k$, we refer to page 281, \cite{CHLL}. However in many cases of $q$ and $n$, there is a linear $[n, k]_q$ code with the covering radius $n-k-1$, see \cite{BGP}. We have the following result.\\

{\bf Corollary 4.2.} {\em Let $q=p^h$ be a prime power with $p \geq 3$, $m$ be a prime factor of $q-1$ satisfying $\max \{7, \frac{(n-k-2)^2}{2}, \frac{3(n-k-2)^2-n+k-12}{2}\} \leq m \leq \frac{1}{8}q^{\frac{1}{4}}$, let $n$ and $k$ be two positive integers satisfying $$\lfloor\frac{q-2\sqrt{q}+1}{m}\rfloor \leq n \leq (\lceil \frac{n-k}{2} \rceil -1)(\lfloor\frac{q-2\sqrt{q}+1}{m}\rfloor+30) +\frac{2(m+1)}{n-k-2}+n-k-\frac{7}{2}.$$ Then an $(n-k-1, L)$ list decodable code ${\bf C} \subset {\bf F}_q^n$ satisfies $$|{\bf C}| \leq L q^{k}.$$}\\

Since $k=n-d_{list}-1$ in the above upper bound, it is stronger than the generalized Singleton bound in \cite{ShangguanTamo} when $L \geq d_{list}=n-k-1$.\\

From the known linear perfect codes such as linear $[\frac{q^m-1}{q-1}, \frac{q^m-1}{q-1}-m, 3]_q$ Hamming perfect code, binary Golay $[23,12,7]_2$ perfect code and ternary Golay $[11, 6, 5]_3$ perfect code, see, Chapter 11, \cite{CHLL}, we have the following result.\\

{\bf Corollary 4.3.} {\em 1) Let $n=\frac{q^m-1}{q-1}$, where $q$ is a fixed prime power and $m=2,3,4, \ldots$. Let $k >n-m$ be a positive integer. Then if a linear $[n,k]_q$ code  is $(1, L)$ list-decodable, then the list size has to satisfy $L \geq q^{k-n+m}$.\\
2) If a binary linear $[23,k]_2$ code is $(3, L)$ list-decodable, then the list size has to satisfy $L \geq 2^{k-12}$.\\
3) If a ternary linear $[11, k]_3$ code is $(2, L)$ list-decodable, then the list size have so satisfy $L \geq 3^{k-6}$.}\\

Our covering code upper bound is much stronger than the generalized Singleton upper bound in \cite{ShangguanTamo} when the lengths are large. It is well-known that the covering radius of the first order $[2^m, m+1, 2^{m-1}]_2$ Reed-Muller code is $2^{m-1}-2^{\frac{m-2}{2}}$ when $m$ is even, see page 243 of \cite{CHLL}. Then the following upper bound for binary $(rn, L)$ list-decodable codes, where $\frac{1}{2}-\frac{1}{2^{\frac{m+2}{2}}} \leq r <\frac{1}{2}$, follows from our covering code bound. The second conclusion follows from Theorem 11.5.3 in page 450 of \cite{HP}. The third conclusion follows from covering radius upper bound for dual BCH codes in \cite{Bazzi}.\\

{\bf Corollary 4.4.} {\em 1) Let $n=2^m$ and $m=2,4,6, \ldots$. Let $r$ be a positive real number satisfying $\frac{1}{2}-\frac{1}{2^{\frac{m+2}{2}}} \leq r <\frac{1}{2}$. If a  binary length $n=2^m$ code ${\bf C}$ is $(rn, L)$ list-decodable, then $$|{\bf C}| \leq L 2^{m+1}. $$ Hence a length $2^m$ binary linear code with the minimum Hamming weight $2^m-2^{\frac{m}{2}}+1$ has its dimension at most $m+1$. \\
2) Moreover for any binary $(s-1, L)$ list-decodable code ${\bf C}$ in ${\bf F}_2^{2s+7}$ where $s$ is any positive integer, we have $|{\bf C}| \leq 64L$. \\
3) Thirdly for code length $n=2^m-1$,  then the cardinality of an $((\frac{1}{2}-(1-o(1))\sqrt{\frac{slog_2n}{n}})n, L)$ list-decodable code ${\bf C} \subset {\bf F}_2^n$, where $s\leq 2^{\frac{m}{2}-1}$, satisfies $$|{\bf C}| \leq L 2^{sm}.$$}\\

For the positive list size $L\geq 2$ we have $\frac{L+1}{L} \leq \frac{3}{2}$, hence the generalized Singleton upper bound in \cite{ShangguanTamo} for this case is at least $$L 2^{2^m -3\cdot 2^{m-2}}=L 2^{2^{m-2}}.$$ Our upper bound is an exponential $2^{\frac{n}{4}-log n+1}$  improvement of the bound in \cite{ShangguanTamo} for binary codes.\\

For a code family of length $n=2^m$, $m=2,4,6,\ldots$ with the rate $R>0$, then if it is $((\frac{1}{2}-\frac{1}{2^{\frac{m+2}{2}}})n, L)$ list-decodable, then the list size $L$ has to be exponential. If we use the list-decoding capacity theorem, $R$ has to bigger than $1-H(\frac{1}{2}-\frac{1}{2^{\frac{m+2}{2}}})$. Hence the conclusion in Corollary 4.4 is a little stronger than the second half of the list-decoding capacity theorem in the binary case. The ratio of the list-decodable radius is of the form $\frac{1}{2}-\frac{c}{\sqrt{n}}$ or $\frac{1}{2}-4.5$ in Corollary 4.4. If we look at the first half of the list-decoding capacity theorem, $1-H(\frac{1}{2}-\frac{1}{2^{\frac{m+2}{2}}})$ is going to zero.  Corollary 4.4 can be compared with Theorem 15 in \cite{GR11}, where the ratio of the list-decodable radius is of the form $$\frac{1}{2}-c\sqrt{\frac{logn}{n}},$$ $c$ is a positive constant. Actually from the results on covering radius upper bounds in \cite{Bazzi} for linear binary codes we get stronger results. For example from Corollary 9 in \cite{Bazzi} about $\frac{2s}{n}^{\frac{2s-5}{13}}$-covering (see \cite{Blin,Blin1}) radius upper bound for dual BCH codes, a stronger upper bound than Corollary 4.4 can be obtained.  This illustrates that our covering code upper bound is strong to translate covering radius upper bounds to list-decodable code upper bounds.\\

Over an arbitrary finite field ${\bf F}_q$ by using the first order $q$-ary Reed-Muller code we can also give an exponential improvement on the generalized Singleton bound in \cite{ShangguanTamo} when the list-decodable codes are very long. From the main result of \cite{Le} the covering radius $R_{covering}(1,m, q)$ of the first order $q$-ary Reed-Muller code is upper bounded by $$R_{covering}(1,m,q) \leq (q-1)q^{m-1}-q^{\frac{m}{2}-1}.$$ Hence we have the following result.\\

{\bf Corollary 4.5.} {\em Let $q$ be an arbitrary prime power. Let $n=q^m$ and $m=1,2,3, \ldots$. Let $r$ be a positive real number satisfying $\frac{q-1}{q}-\frac{1}{q^{\frac{m+2}{2}}} \leq r <\frac{q-1}{q}$. If a  $q$-ary code ${\bf C} \subset {\bf F}_q^n$ is $(rn, L)$ list-decodable, then $$|{\bf C}| \leq L q^{m+1}.$$}\\

This upper bound can be compared with the bound in \cite{GV05,GV10}. The $\epsilon_n$ in Corollary 4.5 is tending to the zero, not a positive constant as in \cite{GV10}. When the list size is at least $L \geq q\sqrt{n}$, the generalized Singleton upper bound in \cite{ShangguanTamo} in this case is at least $q^{n-\frac{1}{q}n}$. In this case our bound is an exponential improvement. This proves that any $((\frac{q-1}{q}-\frac{1}{\sqrt{n}})n, L)$ list-decodable code for $L \leq poly(n)$, when $n=q^m$, has its rate at most $\frac{log_q n+1}{n}$.\\

We can use the previous known result about the covering radius of the $\frac{m}{2}$-th length $2^m$ Reed-Muller code, when $m$ is even, see Chapter 9 of \cite{CHLL} to get the further upper bound as follows. From the result in page 257 of  \cite{CHLL}, the covering radius satisfies $R_{covering}(\frac{m}{2},m) \leq 2^{m-2}-2^{\frac{m}{2}}+2$. Then the following result follows from our covering code upper bounds.\\

{\bf Corollary 4.6.} {\em Let $n=2^m$ and $m=2,4,6, \ldots$. Let $r$ be a positive real number satisfying $\frac{1}{4}-\frac{1}{2^{\frac{m}{2}}} \leq r <\frac{1}{4}$. If a  binary length $n=2^m$ code ${\bf C}$ is $(rn, L)$ list-decodable, then $$|{\bf C}| \leq L 2^{\Sigma_{i=0}^{\frac{m}{2}}\displaystyle{m \choose i}}.$$}\\

Notice that this upper bound can be written as  $L 2^{n-\frac{n-\displaystyle{m \choose \frac{m}{2}}}{2}}$. \\

For give two positive integers $k$ and $d$, let $n_q(k,d)$ be the minimum length $n$ such that there is a linear $[n,k,d]_q$ code. Then the covering radius $R$ of this linear $[n_q(k,d),k,d]_q$ code satisfies $$R_{covering}(n_q(k,d)) \leq d-\lceil \frac{d}{q^k} \rceil,$$ see Corollary 8.1 in \cite{Janwa1}. Then the following result follows immediately from our covering code upper bounds.\\

{\bf Corollary 4.7.} {\em Given two positive integers $k$ and $d$ and $n_q(k,d)$ defined as above. Then for an $(d-\lceil \frac{d}{q^k}\rceil, L$ list decodable code ${\bf C} \subset {\bf F}_q^{n_q(k,d)}$, we have $$|{\bf C}| \leq L q^k.$$} \\

For example for $k=4$ and $d=12$, $n_2(4,12)=23$. An $(12, L)$ list-decodable code ${\bf C} \subset {\bf F}_2^{23}$ has its size $|{\bf C}| \leq 16L$. This is better than the upper bound from \cite{ShangguanTamo} for list size $L \geq 2$ again.\\

Let $r$ and $R$ be two fixed positive integers satisfying $r\geq R$. Let $l_q(r,R)$ be the smallest length of linear of a $q$-ary code with the covering radius $R$ and the redundancy $r=n-k$.  It is clear that there are  $[n, n-r]_q$  code with the covering radius $R$ for each integer $n \geq l_q(r,R)$. Actually this can be proved directly from the the characterization of  the covering radius of linear codes from its parity check matrix. There are a lot of results about this topic, we refer to \cite{DMP1,DMP,BDMP} and references therein.  From our covering code upper bounds the size of an $(R, L)$ list-decodable code ${\bf C}$ of length $n$ satisfying $n \geq l_q(r,R)$ has to satisfy $$|{\bf C}| \leq L q^{n-r}.$$ When the list size is not one, and $r \geq \frac{3}{2}R$, the upper bound is better than the generalized Singleton bound in \cite{ShangguanTamo} . For example for $r=5>\frac{3}{2} \cdot 3$, the size of a length $n \geq l_q(5,3)$ $(3,L)$ list-decodable code satisfies $|{\bf C}| \leq L q^{n-5}$.  In this case $l_q(5,3) < 2.884q^{\frac{2}{3}} (ln q)^{\frac{1}{3}}$ was proved, see \cite{BDMP}. More generally the following result follows from the bound in \cite{DMP}.\\

{\bf Corollary 4.8.} {\em Let $q \geq 8$ be an even prime power. Let $R \geq 4$ be a positive integer. Set $m=\lceil log_q(R+1) \rceil +1$. For any given positive integer $t \geq 3m+2 $, and the length $n \geq Rq^{(t-1)R}+2q^{t-2}+\Sigma_{j=3}^{m+2} q^{t-j}$, the size of an $(R, L)$ list-decodable code ${\bf C} \subset {\bf F}_q^n$ has to satisfy $$|{\bf C}| \leq L q^{n-tR}.$$}

{\bf Proof.} The conclusion follows from Theorem 8 in \cite{DMP}.\\

Comparing with the generalized Singleton bound in \cite{ShangguanTamo}, the upper bound is $|{\bf C}| \leq L q^{n-\lfloor\frac{L+1}{L}R\rfloor}$ is much weaker. Hence when the code length is large the bound in \cite{ShangguanTamo} is far away from the tight.  If we consider the case of $L=1$, Corollary 3.8 implies that when $n \geq Rq^{(t-1)R}+2q^{t-2}+\Sigma_{j=3}^{m+2} q^{t-j}$, the size of a length $n$ code with the Hamming distance $2R+1$ has at most $q^{n-tR}$ codewords. Thus a length $n$ linear code with the minimum Hamming weight $2R+1$ has its dimension $k \leq n-tR$. This is a nontrivial upper bound on the size of codes with the given minimum Hamming weight.\\

The following result gives a lower bound for $L_2-L_1$ for an arbitrary code as an $(R_{list}, L_1, L_2)$ covering list-decodable code.\\

{\bf Corollary 4.9.} {\em Let ${\bf C} \subset {\bf F}_q^n$ be an $(R_{list}, L_1, L_2)$ covering list-decodable code. Suppose that ${\bf C}' \subset {\bf F}_q^n$ is a covering code with the covering radius $R_{list}$ and ${\bf C}'' \subset {\bf F}_q^n$ is a code with the minimum Hamming distance $2R_{list}+1$. Then $$L_2-L_1 \geq |{\bf C}|(\frac{1}{|{\bf C}'|}-\frac{1}{|{\bf C}''|}).$$ Let $A_q(n,2R_{list}+1)$ be the size of the maximal code with the Hamming distance $2R_{list}+1$. Then we have $$L_1A_q(n,2R_{list}+1) \leq |{\bf C}| \leq L_2K_q(n, R_{List}).$$ Hence for a given length $n$ code as a $(R_{list}, L_1, L_2)$ covering list-decodable code, then $$L_2-L_1 \geq |{\bf C}|(\frac{1}{K_q(n, R_{list})}-\frac{1}{A_q(n, 2R_{list}+1)}).$$}\\

\subsection{Covering code dual upper bounds for list-decodable codes}

One existing point of the classical theory of covering codes is the upper bounds on the covering radius of a code from its dual codes. This was originated from the work of \cite{Delsarte,Tie1,Tie}.  We refer to \cite{Delsarte,Tie1,Tie,AHLL,Bazzi,Sole} for some upper bounds on the covering radius from  dual code. The Delsarte upper bound on the covering radius of a linear code ${\bf C}$ asserts that $$R_{covering}({\bf C}) \leq s,$$ where $s$ is the number of nonzero weights of its dual code. Hence we have the following result.\\

{\bf Corollary 4.10 (Delsarte type bound).} {\em  Let ${\bf C} \subset {\bf F}_q^n$ be an  $(d_{list}, L)$ list-decodable code. Suppose that ${\bf C}_1 \subset {\bf F}_q^n$ is a linear code such that the total number of nonzero weights of ${\bf C}_1^{\perp}$ is smaller than or equal to $d_{list}$, then we have $$|{\bf C}| \leq L|{\bf C}_1|.$$}\\

We refer the Delsarte upper bound on the covering radius of linear codes to \cite{Delsarte,Tie,Janwa}.  There are many constructions of linear codes with few nonzero weights, for example we refer to \cite{Ding}.  From the Delsarte upper bound on the covering radius, we have a lot of covering code upper bounds for $(d_{list}, L)$ list-decodable codes for small $d_{list}$. Let $q=p^m$ be an odd prime power with the even exponent $m$, then a linear $[\frac{q-1}{2}, m, \frac{(p-1)(q-\sqrt{q})}{2p}]_p$ code with two nonzero weights is explicitly given in Corollary 4 of \cite{Ding}. Hence for any $(2,L)$ list-decodable code ${\bf C} \subset{\bf F}_p^{\frac{q-1}{2}}$, we have $$|{\bf C}| \leq L p^{\frac{q-1}{2}-m},$$ from the above Delsarte type  upper bound. This is again much stronger than the generalized Singleton upper bound in \cite{ShangguanTamo}. \\

From Theorem 1 of \cite{XLZD} we get the following upper bound from the above Delsarte type upper bound on length $q^2-1$ code over ${\bf F}_p$, $q=p^m$. $h$ is a fixed positive integer satisfying $h \neq 0$ $mod$ $q+1$, $e=\gcd(h,q+1)$, $t \leq \frac{q+1}{2e}$. Then when $p=2$ the cardinality of an $(2t+1, L)$ list-decodable code ${\bf C} \subset {\bf F}_2^{2^{2m}-1}$ has to satisfy $$|{\bf C}| \leq L \cdot 2^{2^{2m}-1-(2t+1)m}.$$ When $p$ is an odd prime, then the cardinality of an $(2t, L)$ list-decodable code ${\bf C} \subset {\bf F}_p^{p^{2m}-1}$ has to satisfies $$|{\bf C}| \leq L \cdot p^{p^{2m}-1-2tm}.$$

Set $\displaystyle{x \choose j}=\frac{(x(x-1) \cdots (x-j+1)}{j!}$, the Krawchouk polynomial is defined by $$K_k(x,q,n)=\Sigma_{j=0}^k \displaystyle{x \choose j} \displaystyle{n-x \choose k-j} (q-1)^{k-j}. $$ For example $K_2(x,q,n)=\frac{1}{2}(q^2x^2-q(2qn-q-2n+2)x+(q-1)^2n(n-1))$.  Let $x(k, q, n)$ be the smallest positive root of $K_k(x,q,n)$.  For a length $n$ linear code with the dual distance $d^{\perp}$ the upper bound for the covering radius is as follows, $R_{covering}({\bf C})  \leq x(u,q,n-1)$ if $d^{\perp}=2u-1$, and $R_{covering}({\bf C})  \leq x(u,q,n)$ if $d^{\perp}=2u$, see \cite{Tie}. For the roots of Krawchouk polynomials we refer to \cite{L,LS}. For example when $q=2$ and $2 \leq u \leq \frac{n}{2}$, $$x(u,2,n) \leq \frac{n}{2}-\sqrt{(n-u+2)(u-2)}.$$

{\bf Corollary 4.11 (Tiet\"{a}v\"{a}inen type bound).} {\em Suppose that ${\bf C}_1$ is a linear $[n,k]_q$ code with the dual distance $d^{\perp}=2u$. Let $d_{list} \geq x(u,q,n)$ be a positive integer. Then the size of an $(d_{list}, L)$ list-decodable code ${\bf C} \subset {\bf F}_q^n$ satisfies $|{\bf C}| \leq L q^k$. In particular a length $n$ code with the Hamming distance $2x(u,q,n)+1$ has its size at most $q^k$.}\\

{\bf Corollary 4.12.} {\em For a given list-decodable radius $d_{list}$ and the code length $n$,  let $u$ be the smallest positive integer such that $x(u,q,n) \leq d_{list}$. Then for the  fixed code length $n$ and the given list-decodable radius $d_{list}$, any $(d_{list}, L)$ list-decodable code ${\bf C} \subset {\bf F}_q^n$ has its cardinality $$|{\bf C}| \leq L \cdot q^{n-\lfloor log_q(\frac{q^n}{\Sigma_{j=0}^{2u-1}\displaystyle{n \choose j} (q-1)^j}) \rfloor}.$$ The cardinality of any $(d_{list}, L)$ list-decodable code ${\bf C} \subset {\bf F}_{q^2}^n$, where $(q^2-1)q^m \leq n \leq (q^2-1)q^{m-1}+2q$ satisfies $$|{\bf C}|\leq L \cdot q^{2u+\frac{n}{q-1}+\frac{n}{q^{\frac{m}{2}-2}}}.$$}\\

{\bf Proof.} These two upper bounds follow from the Tiet\"{a}v\"{a}inen type upper bound, the Gilbert-Varshamov bound or the Drinfeld-Vl\'{a}dut bound of algebraic curves , see \cite{Garcia,TV} directly.\\

\subsection{Sizes of codes with the given minimum Hamming distances}

Even in the case of $(\lfloor\frac{d-1}{2}\rfloor, 1)$ list-decodable case, our covering code bound gives highly nontrivial upper bound on the sizes of codes of the given minimum distance as follows.\\

{\bf Corollary 4.13.} {\em For given minimum distance $d$ and the code length $n$,  let $u$ be the smallest positive integer such that $x(u,q,n) \leq \frac{d-1}{2}$. Then for fixed code length $n$ and the Hamming distance $d$, any linear $[n, k, d]_q$ code has its dimension $$k \leq n-\lfloor log_q(\frac{q^n}{\Sigma_{j=0}^{2u-1}\displaystyle{n \choose j} (q-1)^j}) \rfloor.$$ The dimension of  any linear $[n, k, d]_{q^2}$ code ${\bf C} \subset {\bf F}_{q^2}^n$ where $(q^2-1)q^m\leq n \leq (q^2-1)q^m+2q$, satisfies $$k \leq 2u+\frac{n}{q-1}+\frac{n}{q^{\frac{m}{2}-2}}.$$}\\

{\bf Corollary 4.14.} {\em Given two given positive integers $d$ and $n$ satisfying $d<n$, Let $u$ be the smallest positive integer in the range $2 \leq u \leq \frac{n}{2}$ satisfying $\frac{n-d+1}{2} \leq \sqrt{(n-u+2)(u-2)}$, then the size of a length $n$ binary code ${\bf C} \subset {\bf F}_2^n$ with the given minimum Hamming distance $d$ satisfies $$|{\bf C}| \leq 2(\Sigma_{j=0}^{2u-1}\displaystyle{n \choose j} ).$$}\\

This result can be compared with the first McEliece-Rodemich-Rumsey-Welch bound in \cite{MRRW,HP}, Corollary 2.7 in \cite{LS} and \cite{NS}. Our approach is direct from our simple covering code upper bound and the Tiet\"{a}v\"{a}inen upper bound. The asymptotic form of Corollary 4.14 is a little weaker than the classical McEliece-Rodemich-Rumsey-Welch bound. However Corollary 4.13 and 4.14 are for finite length linear codes.\\

\subsection{Asymptotic bounds}

Based on the asymptotic bound $k_n(q, \rho) \leq 1-H_q(\rho)+O(\frac{log n}{n})$, see \cite{CF}, and the above covering code upper bound, it follows immediately that if the code family with the rate $R \geq 1-H_q(\rho)+\epsilon$,  is $(\rho n,  L)$ list-decodable then the list size $L$ has to be the exponential. This recovers the classical result due to Elias and Zyablov-Pinsker, see Theorem 2.3, page 18 of \cite{Rudra}. This is the second half of the list-decoding capacity theorem.\\

The definition of $U_q^{const}(R)$ and $U_q^{poly}(R)$ are similar to the definitions in \cite{GHSZ} for binary codes. We have $$R \leq 1-H_q(U_q^{const}(R)),$$ and $$R \leq 1-H_q(U_q^{poly}(R)),$$ from the asymptotic form of our covering code bound of Theorem 3.1. Hence if $U_q^{const}(R)$ and $U_q^{poly}(R)$ are in the range $(0, \frac{q-1}{q})$, we have $$U_q^{const}(R) \leq H_q^{-1}(1-R),$$ and $$U_q^{poly}(R) \leq H_q^{-1}(1-R).$$  This partially recovers the bound in \cite{B}, see Theorem 6 in \cite{GHSZ}.\\

On the other hand for the given relative Hamming distance $\delta$ from  Gilbert-Varshamov bound there exists a code family with the relative Hamming distance $\delta$ and rate $R\geq 1-H_q(\delta)$. From the covering code upper bound we have $$1-H_q(\delta) \leq 1-H_q(L_q^{const}(\delta)),$$ and $$1-H_q(\delta) \leq 1-H_q(L_q^{poly}(\delta)).$$

{\bf Corollary 4.15 (asymptotic bound from covering codes).} {\em  If $U_q^{const}(R)$ and $U_q^{poly}(R)$ are in the range $(0, \frac{q-1}{q})$, we  have $U_q^{const}(R)  \leq H_q^{-1}(1-R),$ and $U_q^{poly}(R)  \leq H_q^{-1}(1-R)$. If $\delta$ and $L_q^{const}(\delta)$ (resp. $L_q^{poly}(\delta)$) are in the range $(0, \frac{q-1}{q})$, then $L_q^{const}(\delta) \leq \delta$ and $L_q^{poly}(\delta) \leq \delta$.}\\

We have the following asymptotic upper bound for the polynomial list size list-decodable binary codes from Corollary 4.12, which is similar to and weaker than the first McEliece-Rudemich-Rumsey-Welch bound. To the best of knowledge, this is the first such kind of asymptotic bound for list-decodable binary codes.\\

{\bf Corollary 4.16.} {\em 1) Let $\delta <\frac{1}{2}$ be a positive real number. Then the rate of an $(\delta n, L(n))$ list-decodable binary code family with the list size $L(n) \leq poly(n)$, when $n$ goes to the infinity, has to satisfy $R(\delta) \leq H(2x)$, where $x<\frac{1}{4}$ is the smallest positive real number satisfying $\frac{1-\delta}{2} \leq \sqrt{(1-x)x}$. \\
2) For any $(\delta n,L(n))$ list-decodable code family over a finite field ${\bf F}_q$ satisfying $q\geq 49$ with the polynomial list size $L(n) \leq poly(n)$, let $u<1$ be the smallest real number satisfying $x(un,q,n) \leq \delta n$, then the rate of this $(\delta n,L(n))$ list-decodable code family satisfies $$R(\delta) \leq 2u+\frac{1}{\sqrt{q}-1}$$.}\\

\subsection{List sizes}

We can use the cover code upper bound to lower bound the list size by the size of an $(R, L)$ list decodable code in ${\bf C} \subset {\bf F}_q^n$. Let ${\bf C}_1 \subset  {\bf F}_q^n$ be a covering code with the radius $R_{covering}({\bf C}_1) \leq R$ with the smallest possible size, then the list size is at list $$\frac{|{\bf C}|}{|{\bf C}_1|} \leq L.$$ The asymptotic case of this argument implies the second half of the list-decoding theorem.\\

We consider the beyond Johnson radius case $d_{linst} =  n-\sqrt{\frac{n(n-d)}{1+\epsilon}}$, where $\epsilon$ is a small positive real number.\\

{\bf Corollary 4.17.} {\em If a linear $[n, k, d]_q$ code over ${\bf F}_q$ satisfying $$n-d \leq (1+\epsilon)\frac{k^2}{n}-c,$$ where $c$ is a small positive real number, is $(n-\sqrt{\frac{n(n-d)}{1+\epsilon}}, L)$ list-decodable,  the list size satisfies $L \geq q^{\frac{cn}{2}}$, which is exponential in the code length $n$.}\\

{\bf Proof.} We take any linear $[n, k']_q$ code as ${\bf C}'$ in Theorem 4.1 satisfying $k'=\sqrt{\frac{n(n-d)}{1+\epsilon}}$, then the covering radius of this code is not bigger than $n-k¡®=(n-\sqrt{\frac{n(n-d)}{1+\epsilon}}$ from the redundancy bound. Hence the conclusion follows.\\

From the covering code upper bounds we have the following result directly.\\

{\bf Corollary 4.18.} {\em Let ${\bf C}$ be a length $n$ code over ${\bf F}_q$ of the covering radius $R_{covering}$. Then it is an $(R_{covering}, L_1 \geq 1, L_2)$ covering list-decodable code satisfying $L_2 \geq \frac{|{\bf C}|}{K_q(n, R_{covering})}.$}\\

\subsection{A generalized Singleton upper bound for average-radius list-decodable codes}

A code ${\bf C} \subset {\bf F}_q^n$ is $(d_{list}, L)$-average-radius list decodable if for any subset $\Lambda \subset {\bf C}$ of size $L$ and any fixed ${\bf x} \in {\bf F}_q^n$, then $$\frac{1}{L}(\Sigma_{{\bf c} \in \Lambda} d_H({\bf x}, {\bf c})) \geq d_{list}.$$ Then it is $(d_{list}, L-1)$ combinatorial list-decodable. Otherwise we have at least $L$ codewords in a Hamming ball with the radius $d_{list}$. The above inequality is not valid. We refer to \cite{GN,RW18,GLMRSW20} for recent results on average-radius list-decodable codes.\\

Hence from our covering code upper bound $$|{\bf C}| \leq (L-1)|{\bf C}_{covering}|,$$ for any covering code ${\bf C}_{covering} \subset {\bf F}_q^n$ with the radius $R_{covering} \leq d_{list}$. The above upper bounds in Corollary 4.1-4.11 are valid for average-radius list-decodable codes. We prove a stronger upper bounds for average-radius list decodable codes.\\

{\bf Theorem 4.2 (Generalized Singleton upper bound for average-radius list-decodable codes).} {\em Let $q$ be a prime power and $n$ be a positive integer satisfying $n \leq q$. Let ${\bf C} \subset {\bf F}_q^n$ be a $(d_{list}, L)$-average-radius list-decodable code then we have $|{\bf C}| \leq \max\{(L-1) (q^{n-d_{list}}-\displaystyle{n \choose d_{list}+1})+(\displaystyle{n \choose d_{list}-1}(q-1)^{d_{list}-1}+\displaystyle{n \choose d_{list}}(q-1)^{d_{list}})\displaystyle{n \choose d_{list}+1}, (L-2)q^{n-d_{list}}\}.$}\\

{\bf Proof.}  Since $n \leq q$ we use a Reed-Solomon $[n, n-d_{list}]_q$ code ${\bf C}_{covering}$ as the covering code. Notice that each translation ${\bf y}+{\bf C}_{covering}$ is also a covering code in ${\bf F}_q^n$ for any ${\bf y} \in {\bf F}_q^n$. The covering radius of these translation codes is $n-(n-d_{list})=d_{list}$. If in each Hamming ball centered at codewords of ${\bf y}+{\bf C}_{covering}$ with the radius $d_{list}$, there are at most $L-2$ codewords of ${\bf C}$, then $|{\bf C}| \leq (L-2)q^{n-d_{list}}$. The conclusion follows. Hence we assume that in one such Hamming ball $B({\bf x}, d_{list})$ centered at ${\bf x} \in {\bf y}+{\bf C}_{covering}$ there are exactly $L-1$ codewords of ${\bf C}$. It is well-known that there are $N=\displaystyle{n \choose n-d_{list}-1}=\displaystyle{n \choose d_{list}+1}$ codewords ${\bf x}_1, \ldots, {\bf x}_N$ in ${\bf y}+{\bf C}_{covering}$ satisfying $$d_H({\bf x}, {\bf x}_i)=d_{list}+1.$$ We consider the $N$ Hamming balls centered at ${\bf x}_1, \ldots, {\bf x}_N$ with the radius $d_{list}$. Hence for each codeword ${\bf c}_i \in {\bf C}$ in $B({\bf x}_i, d_{list})-B({\bf x},d_{list})$, and $L-1$ codewords in $B({\bf x}, d_{list}) \bigcap {\bf C}$, we have $$\Sigma_{{\bf c} \in B({\bf x}, d_{list}) \bigcap {\bf C}} d_H({\bf x}, {\bf c})+d_H({\bf x}, {\bf c}_i) \geq Ld_{list}.$$ Since $d_H({\bf x}, {\bf c}_i) \leq d_H({\bf x}_i, {\bf c}_i)+d_H({\bf x}_i, {\bf x})$, then $$\Sigma_{{\bf c} \in B({\bf x}, d_{list}) \bigcap {\bf C}} d_H({\bf x}, {\bf c})+d_H({\bf x}_i, {\bf c}_i) +d_H({\bf x}, {\bf x}_i) \geq Ld_{list}.$$  By the using of a suitable translation we can assume that $d_H({\bf x}, {\bf c})=0$ for one codeword ${\bf c} \in B({\bf x}, d_{list}) \bigcap {\bf C}$, then $d_H({\bf x}_i, {\bf c}_i) \geq d_{list}-1$. That means each codeword ${\bf c}_i \in (B({\bf x}_i, d_{list})-B({\bf x},d_{list})) \bigcap {\bf C}$, has distance at least $d_{list}-1$ to the center ${\bf x}_i$. Then the upper bound follows.\\

This upper bound is stronger than the upper bound $|{\bf C}| \leq (L-1)q^{n-d_{list}}$ when $d_{list}$ is small comparing to $n$ and $log_q L$. Hence this generalized Singleton upper bound for average-radius list-decodable codes can be thought as a distinguishing upper bound from the list-decodable codes.\\

\subsection{Two open problems}

If there was a good upper bound for $K_q(n,d_{list})$ we could lower bound the list sizes of $(d_{list}, L)$ list-decodable codes. The following two problems are natural.\\

{\bf Open problem 1.}\\

Let $q$ be a fixed prime power and $\rho$ be a small positive real number and ${\bf C}_n(\rho n, q) \subset {\bf F}_q^n$ be the covering code of the radius $\rho n$ with the smallest possible size $K_q(\rho n, n)$, when $n$ goes to the infinity, ${\bf C}_n(\rho n, q)$ is a $(\rho n, L_1 \geq 1, L_2)$ combinatorial covering list-decodable code, when can  the second list size $L_2$ be bounded by a constant or $poly(n)$? Or is there a family of covering codes ${\bf C}_n$ of ${\bf F}_q^n$ such that $R_{covering}({\bf C}_n)=\rho n$ and $mul({\bf C}_n) \leq poly(n)$?\\

{\bf Open problem 2.}\\

Can we find $n <q$ special evaluation points in ${\bf F}_q$ such that the corresponding Reed-Solomon $[n,k]_q$ code is a nice $(R_{list}, L_1\geq 1, L_2)$ covering list-decodable code with the smallest possible $L_2-L_1$? Notice that in this case $R_{list} \geq n-k \geq n-\sqrt{n(k-1)}$ is larger than the Johnson radius.\\

\section{List-decodability of rank-metric codes}

List-decoding and list-decodability of rank-metric codes have been studied since 2012, we refer to \cite{WZ,GW1,YDing,RZ}. The rank-metric on the space ${\bf M}_{m \times n}({\bf F}_q)$ of size $m \times n$ matrices over ${\bf F}_q$ is defined by the rank of matrices, i.e., $d_r(A,B)= rank(A-B)$. The minimum rank-distance of a code ${\bf C} \subset {\bf M}_{m \times n}({\bf F}_q)$ is defined as $$d_r({\bf C})=\min_{A\neq B} \{d_r(A,B): A \in {\bf C}, B\in {\bf C} \}$$  The rate of this code ${\bf C}$ is $rate({\bf C})=log_{q^{mn}} |{\bf C}|$. For a code ${\bf C}$ in ${\bf M}_{m \times n}({\bf F}_q)$ with the minimum rank distance $d_r({\bf C}) \geq d$, it is well-known that the number of codewords in ${\bf C}$ is upper bounded by $q^{\max\{m,n\}(\min\{m,n\}-d+1)}$ , see \cite{Gabidulin}. The Gabidulin code is consisting of ${\bf F}_q$ linear mappings on ${\bf F}_q^n \cong {\bf F}_{q^n}$ defined by $q$-polynomials $a_0x+a_1x^q+\cdots+a_ix^{q^i}+\cdots+a_tx^{q^t}$, where $a_t,\ldots,a_0 \in {\bf F}_{q^n}$ are arbitrary elements in ${\bf F}_{q^n}$, is an MRD code, see \cite{Gabidulin}. The rank-distance of is at least $n-t$ since there are at most $q^t$ roots in ${\bf F}_{q^n}$ for each such $q$-polynomial. There are  $q^{n(t+1)}$ such $q$-polynomials. Hence the size of the Gabidulin code is $q^{n(t+1)}$. This is an MRD code. \\

It was shown in \cite{WZ,RZ} that Gabidulin codes cannot be list-decodable to the Johnson radius. The similar result can  also be obtained for rank-metric codes containing Gabidulin codes.  In this section we prove that from the covering code upper bounds, the non-list-decodability of rank-metric codes is rooted in the size of these codes. The following result is for general rank-metric codes, not only Gabidulin codes. Without loss of generality we only restrict to the case $m=n$. Our results essentially rely on the results of Gadouleau-Yan \cite{GY3} on covering rank-metric codes. As in \cite{GY3} let $K_R(q^m,n,\rho)$ be the minimal size of covering rank-metric codes in ${\bf M}_{m \times n}({\bf F}_q)$ with the radius $\rho$, we recall Proposition 11 in \cite{GY3}.\\

{\bf Proposition 5.1 (see Proposition 11 in \cite{GY3}).} {\em Given fixed positive integers $n \leq m$ and $\rho \leq n$, for any $0 < l \leq n$ and $(n_i, \rho_i)$, $i=0, \ldots, l-1$ so that $n_i+\rho_i \leq m$ for all $i$, and $\Sigma_i n_i=n$, $\Sigma \rho_i=\rho$, we have $$K_R(q^m, n, \rho) \leq \min_{(n_i,\rho_i), 0\leq i \leq l-1} \{q^{m(n-\rho)-\Sigma \rho_i(n_i-\rho_i)}\}.$$}\\

{\bf Theorem 5.1.} {\em Let $n$ and $\rho$ be two positive integers satisfying $\rho<n$. Suppose that ${\bf C} \subset {\bf M}_n({\bf F}_q)$ is an $(\rho, L)$-list decodable rank-metric code. Then $$|{\bf C}| \leq L\cdot q^{(n-\rho)(n-\rho+1)-\rho}.$$ Then the size is at least $$L \geq \frac{|{\bf C}|}{q^{(n-\rho)(n-\rho+1)-\rho}}.$$}\\

{\bf Proof.}  We set $m=n$.  From Proposition 3.1,  set $l=2$, $n_0=n-\rho, \rho_0=\rho-1$, $n_1=\rho, \rho_1=1$, we have $$K_R(q^n,n,\rho) \leq q^{(n-\rho)(n-\rho+1)-\rho}.$$ The conclusion follows from Theorem 2.1 1) immediately.\\

For example $K_R(2^6, 4,2)\leq 256$ as in Table I of \cite{GY3}, then any $(2,L)$ list-decodable code ${\bf C} \subset {\bf M}_{6 \times 4}({\bf F}_2)$ has to satisfy $|{\bf C}| \leq 256L$. We can get many such upper bounds on the code sizes and the lower bounds on the list sizes from Table 1 in \cite{GY3}.\\

From Theorem 5.1 we have the following result.\\

{\bf Corollary 5.1.} {\em Let $k, \rho, n$ be three positive integers satisfying $\rho <n, k <n$ and $$k-(n-\rho)(1-\frac{\rho-1}{n})+\frac{\rho}{n} \geq c$$ for a fixed positive real number $c$. If a rank-metric code ${\bf C} \subset {\bf M}_n({\bf F}_q)$ of the cardinality at least $q^{nk}$ is $(\rho, L)$ list-decodable, then the list size is at least $q^{cn}$,  exponential in $n$.}\\

 For a rank-metric code with the rate $\frac{k}{n}>(1-\epsilon)^2$, when $n$ goes to the infinity,  cannot be list-decodable to the radius $\epsilon n$,  where $\epsilon n$ is any positive fixed real number beyond the half minimum distance. Some such kind results for Gabidulin codes were proved in \cite{WZ,RZ}.\\

We consider the asymptotic case. Suppose that $\lim \frac{n}{m}=b$ where $b\in (0,1)$ is a fixed real number. It was shown in \cite{GY1} $$k(r)=\lim_{n \longrightarrow \infty} inf[log_{q^{mn}} K_R(q^m,n,r)] =(1-r)(1-br).$$ Hence any rank-metric $(rn, L)$ list-decodale code family in ${\bf M}_{m \times n}({\bf F}_q)$ has its rate at most $(1-r)(1-br)$  from Theorem 3.1 2). This recovers the first half of the main result in \cite{YDing}.\\

\section{List-decodability of subspace codes}

List-decoding for subspace codes have been studied in \cite{RST,BS,YDing,RZ,MV13,BS,MV19}. The research on subspace codes including constant dimension codes and mixed dimension codes was originated from the paper \cite{KK} of R. K\"{o}tter and F. R. Kschischang. It was proposed to correct errors and erasures in network transmissions of information. A set ${\bf C}$ of $M$ subspaces of the dimension $k \in T$ in ${\bf F}_q^n$, where $T$ is a subset of $\{1,2,\ldots, n-1\}$,  is called an $(n, M, d, T)_q$ subspace code if  $d_S(U,V)=\dim U+\dim V-2\dim(U \cap V) \geq d$ is satisfied for any two different subspaces $U,V$ in ${\bf C}$.  The main problem of the subspace coding is to determine the maximal possible size ${\bf A}_q(n, d, T)$ of such a code for given parameters $n,d,T,q$. When $T$ is the whole set $\{1,2,\ldots,n\}$, we write ${\bf A}_q(n,d)$ for the maximal possible size of the set of subspaces in ${\bf F}_q^n$ such that the subspace distances between any different subspaces in this set are at least $d$. Let $\displaystyle{n \choose k}_q=\prod_{i=0}^{k-1} \frac{q^{n-i}-1}{q^{k-i}-1}$ be the $q$-ary Gauss coefficient, which is the number of $k$-dimensional subspaces in ${\bf F}_q^n$. It is clear ${\bf A}_q(n, d, T) \leq \Sigma_{k \in T}  \displaystyle{n \choose k}_q$ and ${\bf A}_q(n, d) \leq \Sigma_{k=1}^{n-1}  \displaystyle{n \choose k}_q.$ When $T=\{k\}$ contains only one dimension this is a constant dimension subspace code, otherwise it is called a mixed dimension subspace code. There have been some upper and lower bounds for ${\bf A}_q(n,d,k)$. We refer to papers \cite{EtzionVardy,XuChen}.\\

We present some asymptotic results about non-list-decodability of codes in the space $Grass({\bf F}_q^n)$, that is the space of subspaces of ${\bf F}_q^n$ with all dimensions $k \in \{1,2,\ldots,n\}$, endowed with the subspace distance $d_S$ as above. In this seeting the list-decoding of K\"{o}tter-Kschischang codes was given in \cite{MV13}. The rate of a code ${\bf C} \subset Grass({\bf F}_q^n)$ is defined as $\frac{log_q|{\bf C}|}{log_q|Grass({\bf F}_q^n)|}$, see page 2100 in \cite{GY2}.  Notice that the volumes of balls in this finite metric space of the same radius are different when the center subspaces have different dimensions. Set $K_S(q,n,\rho)$ the minimal size of covering codes in this space with the radius $\rho$, and $k_S(r)=\liminf_{n \longrightarrow \infty} \frac{log_qK_S(q, n, rn)}{log_q|Grass({\bf F}_q^n)|}$. It was shown in Proposition 10 of \cite{GY1}, $$k_S(r)=1-2r,$$ for $r \in [0,\frac{1}{2}]$.\\

The injection metric on $Grass({\bf F}_q^n)$ is defined as $d_I(U, V)=\frac{1}{2}(d_S(U,V)+|\dim U-\dim V|$.  In the constant dimension case $d_I=d_S$. In general $\frac{1}{2} d_S\leq d_I \leq d_S$.  Set $K_I(q,n,\rho)$ the minimal size of covering codes in this space endowed with the injection metric,  of the radius $\rho$, and $k_I(r)=\liminf_{n \longrightarrow \infty} \frac{log_qK_I(q, n, rn)}{log_q|Grass({\bf F}_q^n)|}$. It was shown in Proposition 10 of \cite{GY1}, $$k_I(r)=(1-2r)^2,$$ for $r \in [0,\frac{1}{2}]$.\\

We have the following asymptotic results from the above asymptotic results about covering subspace codes under the subspace metric or injection metric.\\

{\bf Theorem 6.1.} {\em 1) If ${\bf C} \subset  (Grass({\bf F}_q^n), d_S)$ is $(\rho, L)$ list-decodable then its size has to satisfy $|{\bf C}| \leq L K_S(q,n,k, \rho)$.  If ${\bf C} \subset  (Grass(k, {\bf F}_q^n), d_I)$ is $(\rho, L)$ list-decodable then its size has to satisfy $|{\bf C}| \leq L K_I(q,n,k, \rho)$.\\
 2) Asymptotically any $(rn, L)$ list-decodale code family in $(Grass({\bf F}_q^n), d_S)$ has its rate at most $1-2r$. Asymptotically any $(rn, L)$ list-decodale code family in $(Grass({\bf F}_q^n), d_I)$ has its rate at most $(1-2r)^2$.}\\

We consider the constant dimension subspace code case. The list-decoding in this setting was studied in \cite{RST,YDing,RZ}, in particular the list-decoding of lifted Gabidulin codes. Let $Grass(k, {\bf F}_q^n)$ be the set of all dimension $k$ subspace of ${\bf F}_q^n$. Then $|Grass(k, {\bf F}_q^n)|=\displaystyle{n \choose k}_q$. Let $K_C(q, n, k, \rho)$  be the minimal size of covering codes in $(Grass(k,{\bf F}_q^n), d_S)$. Several upper bounds on $K_C(q, n, k, \rho)$ were proved in \cite{GY3}. We have the following result.\\

{\bf Proposition 6.1.} {\em If ${\bf C} \subset  Grass(k, {\bf F}_q^n)$ is $(\rho, L)$ list-decodable then its size has to satisfy $|{\bf C}| \leq L K_C(q,n,k, \rho)$. Hence $L \geq \frac{|{\bf C}|}{\displaystyle{n \choose k} \cdot K_R(q^{n-k},k,\rho)}$.}\\

{\bf Proof.} It follows from Proposition 4 of \cite{GY3} immediately.\\

We consider the  case $Grass(n, {\bf F}_q^{2n})$, from Proposition 4.1 we have $|{\bf C}| \leq L q^{(n-\rho)(n-\rho+1)-\rho}\cdot 2^{2n}$. Then the following result follows.\\

{\bf Theorem 6.2.} {\em Let $k,\rho, n$ be three positive integers satisfying $\rho<n, k<n$, and $k-(n-\rho)(1-\frac{\rho-1}{n})+\frac{\rho}{n} \geq c$ for a fixed positive integer $c>2$. If a constant dimension subspace code of the size at least $q^{nk}$ is $(\rho, L)$ list-decodable, then the list size is at least $q^{(c-2)n}$, exponential in $n$.}\\

It follows that some lifted large rank-metric codes in $Grass(n, {\bf F}_q^{2n})$ can not be list-decodable to any positive radius beyond the half minimum distance, when $n$ goes to the infinity. This improves the previous results in \cite{RST,RZ}.\\

\section{Upper bounds on list-decodable insertion codes, deletion codes, cover metric codes and symbol-pair codes}

The covering code upper bound in Theorem 3.1 is general in the following sense that any upper bound on covering codes or covering radius with respect to any metric can be translated to upper bounds on the sizes of list-decodable codes or the lower bounds on the list sizes. However not like the Hamming metric case, there are few results about covering codes in insdel metric, cover metric and pair metric. Hence we can only get weak results in these three cases.\\

\subsection{Generalized Singleton bounds for list-decodable cover metric codes and list-decodable symbol pair codes}

The list-decodability and list-decoding of cover metric codes have been studied in \cite{WZ2,LXY}. The cover metric $d_{cover}$ on the space ${\bf M}_{m \times n}({\bf F}_q)$ of $m \times n $ matrices over ${\bf F}_q$, $m \geq n$,  is defined as follows. Let $A \in {\bf M}_{m \times n}({\bf F}_q)$, a subset pair $(I,J)$,  $I \subset \{1,2\ldots,m\}$, $J \in \{1,2,\ldots,n\}$, is called a cover of $A$ if for any entry of $A$ at the $(i,j)$ position $a_{ij} \neq 0$, then $i \in I$ or $j \in J$.  The cover weight $wt_C(A)$ of $A$ is defined as the minimum $|I|+|J|$ for all such covers. The distance of $A, B \in {\bf M}({\bf F}_q)$ is defined by $$d_{cover}(A,B)=wt_C(A-b).$$ This is a metric on ${\bf M}({\bf F}_q)$, see \cite{WZ2,LXY}.\\

It is obvious there is a covering code in $({\bf M}_{m \times n}({\bf F}_q)$ with the cardinality $q^{mk}$ and the covering radius $n-k$. Hence we have the following generalized Singleton bound on the $(d, L)$ list-decodable codes with the cover metric from Theorem 3.1.\\

{\bf Proposition 7.1.} {\em If ${\bf C} \subset ({\bf M}_{m \times n}({\bf F}_q), d_{cover})$ be an  $(d, L)$ list-decodable code, then we have $$|{\bf C}| \leq L q^{n-d}.$$}\\

This is essentially same as Theorem 1 in \cite{LXY}.\\

The pair metric on ${\bf F}_q^n$ is defined as follows. For ${\bf x}=(x_1, \ldots, x_n) \in {\bf F}_q^n$, we define $pair({\bf x})=((x_0,x_1), (x_1,x_2), \ldots, (x_{n-1},x_n),(x_n,x_0))$. The pair distance $d_{pair}$ is defined by $$d_{pair}({\bf x}, {\bf y})=d_{H}(pair({\bf x}), pair({\bf y})),$$ see \cite{CKWY,LXY1}. The list-decodability and list-decoding of symbol-pair codes have  been studied in \cite{LXY1}.  The following generalized Singleton bound for $(d,L)$ list-decodable codes with the pair metric follows from our general covering code upper bound Theorem 3.1.\\

{\bf Proposition 7.2.} {\em If ${\bf C} \subset ({\bf F}_q^n, d_{pair})$ be an  $(d, L)$ list-decodable code, then we have $$|{\bf C}| \leq L q^{n-d+2}.$$}\\

{\bf Proof.} We construct a linear $[n,k]_q$ covering code with the first $k$ positions as information set positions. Then at the first $k$ positions, for any given vector in ${\bf F}_q^k$, there is codeword whose the first $k$ coordinate vector equal to this vector. Then the covering radius of this code with the pair metric is at most $n-k+2$. The conclusion follows immediately from Theorem 3.1.\\

This recovers the result in \cite{CKWY}, see \cite{CKWY,LXY1}.\\

\subsection{List decodable deletion codes and insertion codes}

It has been a notorious difficult problem to construct efficient codes to correct insertion or deletion errors, see \cite{HS21} for a nice survey. Haeupler and Shahrasbi introduced the concept to synchronization strings and obtained near-Singleton rate-distance tradeoff by using indexing based on their synchronization strings in \cite{HS17,HS18}. The codes constructed in \cite{HS17,HS18} have efficient encoding and efficient unique decoding within the half insdel distances.\\

Let ${\bf A}$ is an alphabet with $v$ elements. The insdel distance $d_{insdel}({\bf a}, {\bf b})$ between two strings ${\bf a} \in {\bf A}^m$ and ${\bf b}$ in ${\bf A}^n$ is the number of insertions and deletions which are needed to transform ${\bf a}$ into ${\bf b}$. Actually $d_{insdel}({\bf a},{\bf b})=m+n-2l$ where $l$ is the length of the longest common substring of ${\bf a}$ and ${\bf b}$. This insdel distance $d_{insdel}$ is indeed a metric. For ${\bf a}$ and ${\bf b}$ in ${\bf A}^n$, it is clear $$d_{insdel}({\bf a}, {\bf b})) \leq 2d_H({\bf a}, {\bf b})$$ since  $l \geq n-d_H({\bf a}, {\bf b})$ is valid for arbitrary two different vectors ${\bf a}$ and ${\bf b}$ in ${\bf F}_q^n$. The insdel distance of a code ${\bf C} \subset {\bf F}_q^n$ is the minimum of the insdel distances of two different codewords in this code. Hence the Singleton upper bound $$|{\bf C}| \leq q^{n-\frac{d_{insdel}}{2}+1}$$  follows from the Singleton bound for codes in the Hamming metric directly,  see \cite{HS17,LTX}. The relative insdel distance is defined as $\delta=\frac{d_{insdel}}{2n}$ since $d_{insdel}$ takes non-negative integers up to $2n$. From the Singleton bound $|{\bf C}| \leq q^{n-\frac{d_{insdel}}{2}+1}$. For insertion-deletion codes the ordering of coordinate positions strongly affects the insdel distances of codes. \\

Efficient binary  list-decodable  insertion-deletion codes have been studied in \cite{HSS18,GHS}. For general bounds on list-decodable insertion-deletion codes, we refer to \cite{WZ1,HY20,HS20,LTX}.  For efficient list-decodable code construction over general fields, see \cite{LTX}.  We refer to \cite{L1,SWY,WY,ASRSU,LRSY21} for the covering codes with insertions or deletions. \\

{\bf Definition 7.1.} {\em A code ${\bf C} \subset {\bf A}^n$ is called $(d, L)$ list-decodable deletion code if for each string ${\bf x}$ in ${\bf A}^{n-d}$,  ${\bf x}$ is a substring of at most $L$ strings in ${\bf C}$. A code ${\bf C} \subset {\bf A}^n$ is called $(d, L)$ list-decodable insertion code if for each string ${\bf x}$ in ${\bf A}^{n+d}$,  ${\bf x}$ contains at most $L$ substrings in ${\bf C}$. A code ${\bf C} \subset {\bf A}^n$ is called $(d_1, d_2, L)$ list-decodable insertion-deletion code, if for each string ${\bf x}$ in ${\bf A}^{n-u+v}$,  where $u \leq  d_2$ and $v \leq d_1$, from arbitrary $u$ insertions and arbitrary $v$ deletions on this string ${\bf x}$,  at most $L$ strings in ${\bf C}$ can be gotten. This condition is hold for any given positive integers $u$ and $v$ satisfying $u \leq d_2$ and $v \leq d_1$.}\\

Notice that from the above definition of $(d, L)$ list-decodable deletion code, each string in ${\bf A}^{n-d'}$, where $d' \leq d$, is substrings of at most $L$ strings in ${\bf C}$. for $(d, L)$ list-decodable insertion code, each string in ${\bf A}^{n+d'}$, where $d' \leq d$, contains at most $L$ substrings in ${\bf C}$. \\

{\bf Definition 7.2.} {\em A code ${\bf C} \subset {\bf A}^n$ is a deletion-covering code of the  radius $d$ if for each string ${\bf x}$ in ${\bf A}^{n-d}$,  ${\bf x}$ is substring of some strings in ${\bf C}$. A code ${\bf C} \subset {\bf A}^n$ is an insertion-covering code of the radius $d$, if for each string ${\bf x}$ in ${\bf A}^{n+d}$,  ${\bf x}$ contains some substrings in ${\bf C}$.  A sequence of codes ${\bf C}_{u, v} \subset {\bf A}^{n-u+v}$, where $u$ and $v$ are any given two positive integers satisfying $u \leq d_2$ and $v \leq d_1$, is called a $(d_1, d_2)$ insertion-deletion covering code sequence of the radius $(d_1,d_2)$, if each string in $ {\bf A}^n$ can be gotten from  $u$ insertions and  $v$ deletions on some string in ${\bf C}_{u,v}$. This condition is hold for any given positive integers $u$ and $v$ satisfying $u \leq d_2$ and $v \leq d_1$. We define the minimum size of this covering code sequence as $$\min_{u \leq d_2, v \leq d_1} \{|{\bf C}_{u,v}|\}.$$  This minimum size is denoted by $K_{insdel}(d_1, d_2, n, q)$.}\\

{\bf Example.} In the binary case ${\bf A}={\bf F}_2$, let ${\bf C}$ be the code consisting of the all-zero string and the all-one string. This is an insertion-covering code in ${\bf C} \in {\bf F}_2^n$ of the radius $n$, since each string in ${\bf F}_2^{2n}$ have a substring of length $n$ with all one or all zero coordinates.  Similarly let ${\bf C} \subset {\bf F}_2^{2n}$ be the code consisting of one string $(0101\ldots 01)$. This is a deletion-covering code of the radius $n$. Then it is direct that we have a cardinality $2^{n-t+1}$ insertion-covering code in ${\bf F}_2^n$ of the radius $t$ and a cardinality $2^{n-2t}$ in deletion-covering code in ${\bf F}_2^n$ of the radius $t$.\\

The following result follows from our general covering code upper bounds Theorem 3.1.\\

{\bf Theorem 7.1.} {\em  1) Let ${\bf C} \subset {\bf A}^n$ be an  $(d, L)$ list-decodable deletion code. Suppose that ${\bf C}_1 \subset {\bf A}^{n-d}$ is a insertion-covering code of  the radius $d$. Then $$|{\bf C}| \leq L|{\bf C}_1|.$$\\
2) Let ${\bf C} \subset {\bf A}^n$ be an  $(d, L)$ list-decodable insertion code. Suppose that ${\bf C}_2\subset {\bf A}^{n+d}$ is a deletion-covering code of the radius $d$.  Then $$|{\bf C}| \leq L|{\bf C}_2|.$$\\
3)  Let ${\bf C} \subset {\bf A}^n$ be an  $(d_1, d_2, L)$ list-decodable insertion-deletion code. Suppose that ${\bf C}_{u,v}\subset {\bf A}^{n-d_1}$, where $u \leq d_2$ and $v \leq d_1$,  is a insertion-deletion covering code sequence of the radius $(d_1,d_2)$. Then $$|{\bf C}| \leq L \min_{u \leq d_2, v \leq d_1} \{|{\bf C}_{u,v}|\}.$$}\\

We have the following result from the previous result in \cite{ASRSU}. Notice that the deletion-covering code in \cite{ASRSU} is the insertion-covering codes in definitions of \cite{L1,SWY,WY,LRSY21}.\\

Corollary 7.1  follows from Theorem 6.14 of \cite{ASRSU}.\\

{\bf Corollary 7.1.} {\em Let  ${\bf C} \subset {\bf A}^n$ be an  $(1, L)$ list-decodable deletion code. Suppose $\frac{n}{log n} \geq 48 v$, then we have $$|{\bf C}| \leq L W(\frac{v^nlog n}{n}),$$ where $W$ is a fixed positive constant.}\\

The following result can be proved from the main results in \cite{WY}.\\

{\bf Corollary 7.2.} {\em Let  ${\bf C} \subset {\bf A}^5$ be an  $(2, L)$ list-decodable insertion code. Suppose that ${\bf C}_1 \subset {\bf A}^7$ is the deletion-covering code of  the radius $2$ determined in \cite{WY}.  Then we have $$|{\bf C}| \leq L|{\bf C}_1|.$$}\\

The following result can be proved from the above example.\\

{\bf Corollary 7.3} {\em If ${\bf C} \subset {\bf F}_2^n$ is an  $(d_1, d_2, L)$ list-decodable insertion-deletion binary code, then we have $|{\bf C}| \leq L \min \{v^{n-2d_2}, v^{n-d_1}\}$.}\\

For insertion-deletion codes, the following problem is natural and important for both covering codes and list-decodable codes.\\

{\bf Open problem 3.} Is there some good upper bound for the minimum size $K_{insdel}(d_1,d_2,n, q)$? Or can we construct some good insertion-deletion covering codes?

\section{List-decodability of sum-rank-metric codes}

Sum-rank-metric codes have applications in network coding, space-time coding and coding for distributed storage, we refer to \cite{MK,MP1,BGR}. For fundamental properties of sum-rank-metric codes, see \cite{MK,MP1,BGR,OPB}. List-decodability of linearized Reed-Solomon codes, which are the codes attaining the Singleton-analogue bound in the sum-rank-metric, have been studied in \cite{PR}. Their results are the generalization of the results in \cite{RZ,WZ} for the rank-metric codes.\\

We recall some basic concepts and results for sum-rank-metric codes in \cite{MK,BGR}. Let $n_i \leq m_i$ be $t$ positive integers satisfying $m_1 \geq m_2 \cdots \geq m_t$. Set $N=n_1+\cdots+n_t$.  Let $${\bf F}_q^{(n_1,m_1), \ldots,(n_t,m_t)}={\bf F}_q^{n_1 \times m_1} \bigoplus \cdots \bigoplus {\bf F}_q^{n_t \times m_t}$$ be the set of all ${\bf x}=({\bf x}_1,\ldots,{\bf x}_t)$, where ${\bf x}_i \in {\bf F}_q^{n_i \times m_i}$, $i=1,\ldots,t$, is the $n_i \times m_i$ matrix over ${\bf F}_q$. Set $wt_{sr}({\bf x}_1, \ldots, {\bf x}_t)=rank({\bf x}_1)+\cdots+rank({\bf x}_t)$ and $$d_{sr}({\bf x},{\bf y})=wt_{sr}({\bf x}-{\bf y}),$$ for ${\bf x}, {\bf y} \in {\bf F}_q^{(n_1,m_1), \ldots,(n_t,m_t)}$. This is indeed a metric on ${\bf F}_q^{(n_1,m_1), \ldots,(n_t,m_t)}$. A code ${\bf C} \subset {\bf F}_q^{(n_1,m_1), \ldots,(n_t,m_t)}$ is a subset and its minimum sum-rank distance is defined by $$d_{sr}=\min_{{\bf x} \neq {\bf y}, {\bf x}, {\bf y} \in {\bf C}} d_{sr}({\bf x}-{\bf y}).$$  The rate of this code is $R=\frac{log_q |{\bf C}|}{\Sigma_{i=1}^t n_im_i}$.\\

The Singleton upper bound for the rank-sum-metric was proved in \cite{MK,BGR}. The general form Theorem III.2 in \cite{BGR} is as follows, Let the minimum sum-rank distance $d$ can be written as the form $d=\Sigma_{i=1}^{j-1} n_i+\delta+1$ where $0 \leq \delta \leq n_j-1$, then $$|{\bf C}| \leq q^{\Sigma_{i=1}^t n_im_i-m_j\delta}.$$ The code attaining this bound is called maximal sum-rank metric distance (MSRD) code. When $m_1=\cdots=m_t=m$ this bound is of the form $$|{\bf C}| \leq q^{m(N-d+1)}.$$\\

The following several special cases of the parameters are important. When $m_1=\cdots=m_t=m$ and $n_1=\cdots=n_t=n$, this is the $t$-sum-rank-metric code over ${\bf F}_{q^m}$ with the code length $N=nt$. When $n=1$, the Hamming metric is recovered. When $t=1$, the rank-metric is recovered. Hence the sum-rank-metric is a generalization and combination of the Hamming metric and the rank-metric.\\

The volume of radius $r$ in the sum-rank-metric is $$vol(B_r({\bf F}_q^{(n_1,m_1), \ldots,(n_t,m_t)}))=\Sigma_{s=0}^r \Sigma_{(s_0,\ldots,s_t): s_0+\cdots+s_t=s} \prod \displaystyle{n_i \choose s_i} \prod_{j=0}^{s_i-1}(q^{m_i}-q^j),$$ we refer to Lemma III.5 in \cite{BGR}. In the case $n_1=\cdots=n_t=n_t$, $m_1=\cdots=m_t=m$, set $$f(z)=\Sigma_{i=0}^n \displaystyle{n \choose i} \prod_{j=0}^{i-1}(q^m-q^j)z^i,$$ and $$H_{sum-rank}(\rho)=\frac{1}{mn}\min_{z \in (0,1]} log_q (\frac{f(z)}{z^{\rho}},$$  where $\rho$ is a positive real number satisfying  $0<\rho<n$. Then from \cite{BRG} Theorem IV.9, when $n,m, \rho <n$ are fixed, $$\lim_{t \longrightarrow \infty} \frac{log_{q^{mn}} (vol(B_{\rho t}(({\bf F}_q^{(n,m), \ldots,(n,m)}))))}{t}=H_{sum-rank}(\rho).$$ From Lemma 2 in \cite{OPB} we have $$H_{sum-rank}(\rho) \geq \frac{m+n-\rho)\rho-\frac{1}{4}-log_q \gamma_q}{mn},$$ where $\gamma_q=\prod_{i=1}^{\infty}(1-q^{-i})^{-1}$, for example $\gamma_2 \approx3.463, \gamma_3 \approx1.785$ and $\gamma_4 \approx1.452.$\\

From our asymptotic covering upper bound in Theorem 3.1 we have the following result.\\

{\bf Theorem 8.1.} {\em When $n<m, \rho<n$ are fixed and $t$ goes to the infinity, if a family of $t$-block sum-rank-metric codes with the rate $R \geq 1-H_{sum-ranl}(\rho)+\epsilon$ is $(\rho t, L)$ list-decodable, then the list size $L$ satisfies $L \geq q^{t\epsilon}$, which is exponential. In particular if a family of sum-rank-metric codes with the rate $R \geq 1-\frac{(m+n-\rho)\rho-\frac{1}{4}-log_q \gamma_q}{mn}+\epsilon$, is $(\rho t, L)$ list-decodable, then the list size $L$ satisfies $L \geq q^{t\epsilon}$, which is exponential.}\\

Set $m=n$ and $d=\lambda mt$, $\rho=\frac{d}{2}$, then when $m \leq \frac{4(1-\lambda)}{\lambda^2}$, the condition in Theorem 5.1 is satisfied. Hence in this case the linearized Reed-Solomon codes satisfying $|{\bf C}|=q^{nt-d+1}$ can not be list-decodable beyond the half-distance. The results can be compared with the main result in \cite{PR}.\\

Now we give a non-list-decodability result generalizing Theorem 3.1 and Corollary 3.1.\\

{\bf Theorem 8.2.} {\em Suppose that $m=n$ fixed and $t$ goes to the infinity. Let $\rho$ be a positive integer satisfying $\rho<n$. Suppose that ${\bf C} \subset {\bf F}_q^{(n,n), \ldots,(n,n)}$ is an $(\rho t, L)$ list-decodable sum-rank-metric code. Then $$|{\bf C}| \leq L\cdot q^{t((n-\rho)(n-\rho+1)-\rho)}.$$ Then the size is at least $$L \geq \frac{|{\bf C}|}{q^{t((n-\rho)(n-\rho+1)-\rho)}}.$$}\\

{\bf Proof.}  Similar to the proof of Theorem 3.1.\\

From Theorem 8.2 we have the following result.\\

{\bf Corollary 8.1.} {\em Let $k, \rho, n$ be three positive integers satisfying $\rho <n, k <n$ and $t$ goes to the infinity. Suppose that $$k-(n-\rho)(1-\frac{\rho-1}{n})+\frac{\rho}{n} \geq c$$ for a fixed positive real number $c$. If a sum-rank-metric code ${\bf C} \subset {\bf F}_q^{(n,n), \ldots,(n,n)}$ of the cardinality at least $q^{nkt}$ is $(\rho t, L)$ list-decodable, then the list size is at least $q^{cnt}$,  exponential in $nt$.}\\

As in the case of rank-metric codes, from our covering code upper bound, the non-list-decodablity of a sum-rank-metric code is rooted in its size. So the subcodes of some linearized Reed-Solomon codes  near the Singleton-analogue bound, not attaining the bound, can not be list-decodable beyond the half minimum distance.\\

When $n_1=n_2=\cdots=n_{t-1}=1$ and $n_t=m_1=\cdots=m_t=n$ are fixed and $t$ goes to the infinity, this case is a combination of the Hamming metric and rank-metric. We consider the list-decodable radius $d_{list}=r(t-1)+\rho $, where $r<1$ is a fixed positive real number and $\rho<n$ is a fixed positive integer. Then the threshold of the non-list-decodablity is $1-H_{q^n}(r)$ as in the Hamming metric case.  However if we allow the cardinality of the base field ${\bf F}_{q^n}$ goes to the infinity, that is $n_t=m_1=\cdots=m=n=\lambda t$, where $\lambda $ is a fixed positive real number, $d_{list}=r(t-1)+\rho t$ where $\rho$ is a fixed positive real number, the code length is $t-1+\lambda t$, the situation is quite different. There is a covering code in the sum-rank-metric of the radius at most $r(t-1)+\rho t$, with the cardinality $K_{q^{\lambda t}}(t-1,r(t-1)) \cdot K_R(q^{\lambda t}, \lambda t, \rho t)$, where $K_{q^{\lambda t}}(t-1,r(t-1))$ is the minimal size of a covering code of radius $r(t-1)$ in the Hamming metric space ${\bf F}_{q^{\lambda t}}^{t-1}$ defined in Section 2, and $K_R(q^{\lambda t}, \lambda t, \rho t)$ is the minimal size of a covering code of radius $\rho t$ in the rank-metric space ${\bf M}_{\lambda t \times \lambda t}({\bf F}_q)$ defined in Section 3. Therefore the threshold of non-list-decodablity is $\frac{1-r+(1-\rho)^2\lambda}{1+\lambda}$, from the previous asymptotic formula in the Hamming metric case and rank-metric case.\\

\section{List-decodable permutation codes with the Hamming metric and the Chebyshev metric}

Let ${\bf S}_n$ be the group of all permutations of $n!$ elements, various metrics are defined on this permutation group and codes with these metrics have been studied for rank modulation, we refer to \cite{BJW72,BCD79,Deza,Cameron,ST2011,FSM13,WMW2015,BEtzion2015,FSB2016,KS2018,Kong,MN2020} for constructions and bounds of permutation codes. In this section we give some non-list-decodablity results for permutation codes with Hamming metric and Chebyshev metric by the using of previous results on permutation covering codes in \cite{Cameron,KS2018,FSB2016}.\\

The Hamming metric on ${\bf S}_n$ is defined as $d_H(\sigma, \tau)=|\{1 \leq i \leq n:\sigma(i) \neq \tau(i)\}|$ and the the Chebyshev metric (infinity norm) is defined by $d_{Ch}(\sigma, \tau)=\max_{1 \leq i \leq n} |\sigma(i)-\tau(i)|$.  For a permutation code ${\bf C} \subset {\bf S}_n$, its minimum Hamming distance is $d_H({\bf C})=\min_{\sigma \neq \tau} d_H(\sigma, \tau)$ and its minimum Chebyshev distance is $d_{Ch}=\min_{\sigma \neq \tau}d_C(\sigma, \tau)$. The rate of a permutation code ${\bf C} \subset {\bf S}_n$ is defined as $R({\bf C})=\frac{log_2 |{\bf C}|}{n}$.\\

The covering radius of the permutation code ${\bf C} \subset {\bf S}_n$ with the Hamming metric is denoted by $R_H({\bf C})$ and the covering radius with the Chebyshev metric is denoted by $R_{Ch}({\bf C})$. For the construction of good permutation codes with these two metrics, see \cite{ST2011,KS2018,MN2020}.  Some upper bounds on the covering radius of some codes with these two metrics have been obtained in \cite{Cameron,KS2018,FSB2016}. Hence we can apply our covering code upper bound directly on list-decodable permutation codes with these two metrics. We refer to \cite{Deza,BEtzion2015,FSB2016,Kong} for permutation codes with the Ulam metric and the Kendall $\tau$-metric.\\

\subsection{Chebyshev metric}

Let ${\bf G}_n \subset {\bf S}_n$ be the subgroup generated by the mapping $(1,2, \ldots, n)$, that is the mapping $f$ satisfying $f(1)=2, f(2)=3, \ldots, f(n-1)=n, f(n)=1$. Then the size of this code is $|{\bf G}_n|=n$. The covering radius of this code ${\bf G}_n$ with the Chebyshev metric was determined in \cite{KS2018}, Theorem 8, $R_{Ch}({\bf G}_n)=n-\lfloor \frac{\sqrt{4n+1}+1}{2} \rfloor$. Hence the following result follows directly from Karni-Schwartz Theorem. The second conclusion follows from the permutation code in page 5224 in \cite{KS2018}.\\

{\bf Theorem 9.1.} {\em  Let ${\bf C} \subset {\bf S}_n$ be an $(n-\lfloor \frac{\sqrt{4n+1}+1}{2} \rfloor, L)$ list-decodable permutation code with the Chebyshev metric, then the cardinality of ${\bf C}$ satisfies $|{\bf C}| \leq n L$. Let $m$ and $t$ be two positive integers. Then an $(m-\lfloor \frac{\sqrt{4m+1}+1}{2} \rfloor, L)$ list-deocodable code ${\bf C} \subset {\bf S}_{mt}$ has its cardinality $$|{\bf C}| \leq L \frac{(mt)!}{((m-1)!)^t}$$.}\\

{\bf Corollary 9.1.} {\em For a family of $(n_i-\lfloor \frac{\sqrt{4n_i+1}+1}{2} \rfloor, L)$ list-decodable permutation codes $\{{\bf C}_i\}$ of the length $n_i$ with the Chebyshev metric satisfying the list size $L \leq poly(n_i)$ when $n_i$ tends to the infinity, the cardinalities of codes satisfy  $|{\bf C}_i| \leq poly(n_i)$ when $n_i$ tends to the infinity.}\\

From the result in \cite{KS2018,WMW2015}, see Lemma 13 in \cite{KS2018},  we have the following asymptotic bound for non-list-decodability of permutation codes with the Chebyshev metric.\\

{\bf Theorem 9.2.} {\em Let $\rho$ be a positive real number satisfying $0<\rho<1$. Set $R(\rho)=-\rho \lfloor\frac{1}{\rho}\rfloor log_2 \rho-(1-\rho \lfloor\frac{1}{\rho} \rfloor)log_2(1-\rho \lfloor\frac{1}{\rho} \rfloor) +o(1)$, where $o(1)$ is a function that tends to zero when $n$ tends to the infinity. Then the rata $R$ of a family of $(\rho n, L)$ list-decodable permutation codes in ${\bf S}_n$ with the Chebyshev metric satisfying $R \leq R(\rho)$, if the list size satisfies $L \leq poly(n)$ when $n$ tends to the infinity.}\\

\subsection{Hamming metric}

Now we consider the permutation covering codes in the Hamming metric. One of the main result in \cite{Cameron} asserts that the minimum cardinality of the covering code ${\bf C} \subset {\bf S}_n$ with the Hamming covering radius $n-s$ satisfying $|{\bf C}| \leq e s! n log_2 n$ provided $n \geq 2s+2$. Then we have the following result for list-decodable permutation codes with the Hamming metric.\\

{\bf Theorem 9.3.} {\em Let $n$ and $s$ be two positive integers satisfying $n \geq 2s+2$. Let ${\bf C}$ be an $(n-s, L)$ list-decodable permutation code with the Hamming metric, then the cardinality of ${\bf C}$ satisfies $|{\bf C}| \leq e L n log_2 n s!$.}\\

The following polynomial size upper bound about polynomial list size list-decodable permutation codes with the list-decodable radius closing to the length follows from Theorem 6.3 immediately. \\

{\bf Corollary 9.2.} {\em Let $s$ be a fixed positive integer. For a family of $(n_i-s, L)$ list-decodable permutation codes $\{{\bf C}_i\}$ of the length $n_i$ with the Hamming metric satisfying the list size $L \leq poly(n_i)$ when $n_i$ tends to the infinity, the cardinalities of codes satisfy  $|{\bf C}_i| \leq poly(n_i)$ when $n_i$ tends to the infinity.}\\

We can compare Corollary 6.1 and 6.2 with the main results in \cite{GV10}. All these results shows that when the list-decodable radius is close to the length, the size of polynomial list size list-decodable codes have their cardinalities bounded by polynomial functions of code lengths.\\

\section{Conclusion}

In this paper we propose to look at $(d_{list}, L)$ list-decodable codes in a general finite metric space $({\bf X},d)$ from various covering codes ${\bf C}'$ of the same length satisfying the covering radius $R_{covering}({\bf C}') \leq d_{list}$. Then any such small ${\bf C}'$ gives a good upper bound on the sizes of $(d, L)$ list-decodable codes in $({\bf X},d)$. The corresponding asymptotic covering code upper bounds lead to some new asymptotic upper bounds for list-decodable codes. This kind of simple upper bounds is a strong constraint on the list-decodability of codes in $({\bf X}, d)$.  Various upper bounds on the sizes of list-decodable codes or the lower bound on the list sizes from various known results about covering codes in the Hamming metric setting are given, to illustrate that if we could find suitable small covering codes with $R_{covering} \leq d_{list}$ the upper bound on the size and the lower bound on the list size can be obtained.  From our general covering code upper bound we also propose a generalized Singleton upper bound for average-radius list-decodable codes, which is stronger than the similar upper bound for list-decodable codes in some parameter range.\\

Our general covering code upper bound can apply to the finite metric space in which the volumes of balls depend not only on radius, but also on centers. Actually from this covering-code view about $(d_{list},L)$ list-decodability, the key ingredient is the size of covering code with the radius $d_{list}$, not the volumes of balls of the radius $d_{list}$. General covering code upper bounds can also be applied to list-decodable rank-metric codes, list-decodable subspace codes,  list-decodable insertion codes, list-decodable deletion codes, list-decodable sum-rank-metric codes and list-decodable permutation codes with various metrics. The covering code in the Hamming metric setting has been a classical topic in coding theory. However there are few results about covering codes with other metrics such as rank-metric, subspace metric, insertion-deletion metric , sum-rank-metric and the Hamming metric, the Chebyshev metric, the Kendall $\tau$-metric and the Ulam metric for permutation space. The basic point of this paper is that upper bounds for covering code sizes or covering radius with various metrics are constraints on list-decodable codes with these metrics. We introduce combinatorial $(R_{list}, L_1, L_2)$ covering list-decodable  codes in a general finite metric space and suggest to study $(R_{list}, L_1=L_2)$ covering list-decodable codes as generalized perfect codes.\\

\end{document}